\documentclass[a4paper,11pt]{article}
\pdfoutput=1 

\usepackage{jcappub} 

\usepackage[T1]{fontenc} 
\usepackage{mathrsfs}
\usepackage{mathtools}
\usepackage[retainorgcmds]{IEEEtrantools}
\usepackage{graphics}
\usepackage{subfigure}
\usepackage{xcolor}
\usepackage{tabularx}
\usepackage{booktabs}
\usepackage{enumerate}

\newcommand{\ud}{\mathrm{d}}

\defcitealias{2017PhRvD..96f3505L}{LSR17}

\title{\boldmath Precision cosmology and the stiff-amplified gravitational-wave background from inflation: NANOGrav, Advanced LIGO-Virgo and the Hubble tension}

\author[a,1]{Bohua Li\note{Corresponding author.}}
\author[b]{and Paul R. Shapiro}

\affiliation[a]{Department of Astronomy, Tsinghua University,\\
Beijing, 100084, China}
\affiliation[b]{Department of Astronomy and Texas Cosmology Center, 
The University of Texas at Austin,\\
2515 Speedway C1400, Austin, Texas 78712, USA}

\emailAdd{bohuali@mail.tsinghua.edu.cn}
\emailAdd{shapiro@astro.as.utexas.edu}

\abstract{
The recent NANOGrav finding of a common-spectrum process has invited interpretations 
as possible evidence of a primordial stochastic gravitational-wave background (SGWB) 
stronger than predicted by standard inflation + $\Lambda$CDM.  
Such an SGWB would contribute an extra radiation component to the background Universe which may affect its expansion history. As such, it may help alleviate the current Hubble tension, a novel connection between gravitational waves and cosmology. 
We demonstrate this by considering a cosmological model, the ``standard inflation + stiff amplification'' scenario, 
with two components added to the base-$\Lambda$CDM model: a stiff component ($w\equiv p/\rho =1$) and the primordial SGWB.
Previously, we showed that even for \emph{standard} inflation, 
the SGWB may be detectable at the high frequencies probed by laser interferometers, 
if it is amplified by a possible early stiff era after reheating. 
Models that boost the SGWB enough to cause significant \emph{backreaction}, however, 
must still preserve the well-measured radiation-matter equality, respecting the demands of precision cosmology. 
For that, we calculate the fully-coupled evolution of the SGWB and expansion history, 
sampling parameter space (tensor-to-scalar ratio, reheating temperature and temperature at stiff-to-radiation equality).
We then perform a joint analysis of the NANOGrav results and latest upper bounds from \emph{Planck}, 
big bang nucleosynthesis and Advanced LIGO-Virgo, to constrain the model.
The resulting blue-tilted, stiff-amplified SGWB is still too small to explain the NANOGrav results. 
However, if someday, Advanced LIGO-Virgo detects the SGWB, 
our model can explain it within standard inflation (\emph{without} requiring an initial spectral tilt).
Meanwhile, this model may bring current high-$z$ measurements of the Hubble constant 
within $3.4\sigma$ of the low-$z$ measurements by SH0ES (from $4.4\sigma$)
and within $2.6\sigma$ of those by H0LiCOW (from $3.1\sigma$), reducing the tension.
}

\begin{document}
\maketitle
\flushbottom

\section{Introduction}
\label{sec:intro}

A stochastic gravitational-wave background (SGWB) from primordial tensor fluctuations 
is generically produced in the inflationary paradigm \citep{1979JETPL..30..682S,1982PhLB..115..189R,1984NuPhB.244..541A}.
Once deemed too small to be detected, this primordial SGWB is now possibly within reach of various experiments, 
from the cosmic microwave background (CMB) to gravitational-wave (GW) interferometers, 
over a wide range of frequencies \citep{2000PhR...331..283M,2016PhRvX...6a1035L}. 
It may even contribute significantly enough to the energy content of the Universe as to affect the expansion history, with possible observable consequences beyond its direct detection \citep[e.g.,][]{1969JETPL...9..184S,2017PhRvD..96f3505L}.  
These direct and indirect probes of the primordial SGWB can, therefore,  
potentially reveal important information about inflation and other physical processes in the early Universe,
which are otherwise poorly understood \citep[e.g.,][]{1992PhR...215..203M,2018JCAP...11..038K}.

In fact, even before inflation was proposed, Grishchuk realized that in an expanding Universe, 
significant \emph{parametric amplification} can occur, not only
for classical gravitational waves (GWs), but even for quantum fluctuations of the vacuum 
\citep{1974ZhETF..67..825G,1977NYASA.302..439G}.
It requires that (1) modes spend time outside the Hubble radius
(i.e., the background Universe expands more rapidly than GWs vary in time), when (2) the Universe is not radiation-dominated (RD).
When both conditions are met, GWs, or tensor fluctuations, will be amplified relative to the ``adiabatic'' solution 
(for which $h\propto 1/a$ for modes always well-inside the Hubble radius).

The inflationary paradigm \citep{1980PhLB...91...99S,1981PhRvD..23..347G,1982PhLB..108..389L}
naturally provides such a period that enables parametric amplification
and production of macroscopic GWs from initial vacuum fluctuations.
When tensor modes are stretched well outside the Hubble radius, 
their amplitudes become time-independent, or ``frozen'' \citep{1979JETPL..30..682S,2003PhRvD..67l3504W}. 
These amplitudes define the primordial tensor power spectrum.
For standard single-field, slow-roll inflation, their distribution is nearly Gaussian, 
with a nearly scale-invariant power spectrum that satisfies the consistency relation \citep{1992PhRvL..69.1856D,1992PhLB..291..391L}.
After inflation ends, tensor modes start to reenter the Hubble radius 
and each, thereafter, evolves according to the adiabatic solution, redshifting like radiation. 
Together, all modes that reentered and remain inside the Hubble radius constitute the primordial SGWB.

The parametric amplification regime for a given mode 
spans its Hubble exit and reentry \citep{1993CQGra..10.2449G}.\footnote{In this paper, 
``Hubble exit/reentry'' refers to the times at which a mode exits/reenters the Hubble volume, 
when its wavelength passes above or below the Hubble radius, respectively.} 
While all modes of interest exit during inflation, 
different modes can reenter during post-inflationary eras with different equations of state (EoS).
This actually leads to another kind of amplification/attenuation of the primordial SGWB, as we describe below, 
in terms of the departure of the amplitudes of modes at a given time after their reentries,  
\emph{relative} to those if the EoS of the Universe during their reentries were radiation-like 
($w\equiv \bar p/\bar\rho=1/3$).

Our observed Universe must undergo a standard RD era
which begins no later than big bang nucleosynthesis (BBN) and ends at radiation-matter equality.
For nearly scale-invariant initial conditions, 
the contribution to the present-day SGWB energy spectrum, $\Omega_{\rm GW}(f)\equiv\ud\,\Omega_{\rm GW}/\ud\ln f$,
by modes that reentered during this RD era is nearly frequency-independent.   
This results in a spectrum with a long ``plateau'' \citep{1997PhRvD..55..435T}. 
In what follows, we shall henceforth use the term \emph{amplification} 
to refer, not to the parametric amplification effect described above,  
but rather to the amplification of a mode at a given time \emph{after} it reenters, 
\emph{relative to this plateau} associated with Hubble reentries that take place during the RD era.  
For modes of longer wavelengths that reenter during the matter-dominated (MD) era ($w=0$) which follows the RD era, 
$\Omega_{\rm GW}(f)$ is amplified relative to the plateau,
since the time-dependence of the Hubble parameter then differs from that of an RD Universe \citep{1989CQGra...6L.161G,1990PhRvD..42.3413G}.
On the other hand, for modes with short enough wavelength to reenter before BBN,
the possibility exists for amplification, too, since the expansion history or, equivalently, 
the EoS of the Universe during this period are poorly constrained and may also depart from $w=1/3$. 
In fact, for these modes of higher frequencies, Giovannini \citep{1998PhRvD..58h3504G}
considered the interesting case in which 
$\Omega_{\rm GW}(f)$ is amplified relative to the plateau by an early phase whose EoS is stiffer than radiation (i.e., $w>1/3$).
This possibility has subsequently been studied by many authors
\citep{1999PhRvD..59f3505P,1999CQGra..16.2905G,1999PhRvD..60l3511G,2008PhLB..668...44G,2008PhRvD..77f3504B,2008PhRvD..78d3531B,2011PhRvD..84l3513K,2017PhRvD..96f3505L,2019JCAP...08..011F}.

In a previous paper \citep[][hereafter LSR17]{2017PhRvD..96f3505L}, 
we investigated this amplification effect in the particular context of complex scalar field dark matter (SFDM)
made up of charged ultralight bosons \citep{2014PhRvD..89h3536L}.
If all cosmological dark matter consists of SFDM (the $\Lambda$SFDM model),
the Universe would be dominated at early times by the stiff phase of SFDM ($w_{\rm SF}=1$), 
before the standard RD era.
The stiff phase of a scalar field is also known as the ``kination'' phase \citep{1997PhRvD..55.1875J}, 
since the energy density of the SFDM is dominated by the kinetic energy.
LSR17 showed that this early stiff-SFDM-dominated era ($w=1$; ``stiff era'' for short) 
indeed amplifies the high-frequency part of $\Omega_{\rm GW}(f)$ relative to the plateau value.
This amplified SGWB may contribute a non-negligible radiation component
to the total energy density, which boosts the expansion rate during the RD era. 
Meanwhile, this same effect results in a blue tilt in $\Omega_{\rm GW}(f)$,
which may even make direct detection of the SGWB possible 
at high frequencies by current laser interferometer experiments, 
e.g., Advanced LIGO-Virgo \citep{2015CQGra..32g4001L, 2015CQGra..32b4001A} and LISA \citep{2017arXiv170200786A}.
Therefore, the stiff-era amplification effect (henceforth, ``stiff amplification'') 
encourages multi-wavelength search of the primordial SGWB using different GW probes
\citep[e.g.,][]{2006PhRvD..73b3504S,2015PhRvD..91j3505M}.
In this paper, we again focus on the stiff-amplified primordial SGWB, in a more general context, not limited to that involving SFDM.

Besides the CMB and laser interferometers, pulsar-timing array (PTA) observations
can probe the SGWB by searching for correlated timing deviations in millisecond pulsars induced by the SGWB
\citep{1979ApJ...234.1100D,2019A&ARv..27....5B}.
Recently, the North American Nanohertz Observatory for Gravitational Waves (NANOGrav) 
reported strong evidence for a stochastic common-spectrum process 
in their 12.5~yr pulsar-timing data set \citep{2020ApJ...905L..34A}
with a high amplitude ($h_c\sim 10^{-15}$ at $f_{\rm yr}=1$~yr$^{-1}$).
Though it has not been confirmed as an SGWB detection yet, 
many interpretations in this direction have flourished since then.
Possible SGWB sources include a cosmic population of supermassive black hole binaries \citep{2021MNRAS.502L..99M},
cosmic strings \citep{2021PhRvL.126d1304E,2021PhRvL.126d1305B,2021PhRvD.103f3031R}, 
phase transitions \citep{2021arXiv210413930A,2021PhLB..81636238N,2020arXiv200910327A,2021ScPP...10...47R},
the primordial SGWB with a large initial blue tilt from non-standard inflationary scenarios (relaxing the consistency relation) \citep{2021MNRAS.502L..11V,2021JCAP...01..071K} 
and others \citep[e.g.,][]{2021PhRvL.126d1303D,2021arXiv210501943Y}.

In this paper, we are, however, interested in the \emph{secondary} blue tilt in the primordial SGWB produced by stiff amplification, 
within the \emph{standard} inflationary scenario which \emph{preserves} the consistency relation
(henceforth, the ``standard inflation + stiff amplification'' scenario).
The case of stiff amplification ($w=1$) is the one that maximizes the possible secondary blue tilt that results for modes that reentered when the EoS of the Universe has $w>1/3$. 
Thus, the first part of this paper is dedicated to the question 
of whether stiff-amplified primordial SGWB can explain the high common-spectrum amplitude reported by NANOGrav.
To this end, we consider a cosmological model with two additional components to the base-$\Lambda$CDM model \citep{2020A&A...641A...6P}: 
a stiff component and the primordial SGWB.
In our model, when inflation ends,  there is an extended phase of reheating with a matter-like EoS ($w=0$) \citep{1982PhRvL..48.1437A,1982PhLB..117...29A}.
When reheating ends, the Universe is assumed to be dominated by the stiff component and remains so until the onset of the RD era.
In order to constrain our model parameters, we perform a joint analysis of the latest observational results 
from the CMB, BBN, NANOGrav and Advanced LIGO-Virgo's third observing run (O3) \citep{2021PhRvD.104b2004A}.

Our analysis has a novel feature:
we \textit{self-consistently} include the \textit{backreaction} of the SGWB on the background expansion rate, 
as we did in LSR17.
Although noted before \citep[e.g.,][]{1973grav.book.....M,1998PhRvD..58h3504G},
this backreaction effect is unfortunately often neglected when modelling the SGWB.
Nevertheless, as stated above, the stiff-amplified SGWB can contribute a non-negligible (percent level) radiation component
to the total energy density during the RD era.
This will, in return, not only affect the evolution of tensor modes, 
but also other observables, e.g., radiation-matter equality and the CMB damping tail.
A \emph{precise} analysis of the primordial SGWB 
ought to account for its coupling with the background expansion history, therefore.

In the meantime, the well-known Hubble tension \citep[e.g.,][]{2016JCAP...10..019B,2017NatAs...1E.169F} 
also motivates our treatment.
The present-day Hubble constant, $H_0$, now measured at better than $3\%$ precision by several experiments, 
shows a discrepancy $(>3\sigma)$ between its value measured by the CMB \citep[][]{2020A&A...641A...6P} 
and that by the distance ladder or time delays of lensed quasars in the nearby Universe \citep{2019ApJ...876...85R,2020MNRAS.498.1420W}.
With respect to the aforementioned radiation-matter equality, 
one way to alleviate the Hubble tension is to exploit the $H_0-N_{\rm eff}$ degeneracy: 
the redshift of this equality can be kept constant 
by increasing the value of $H_0$ and the effective number of relativistic species at the same time \citep[][]{2004PhRvD..69h3002B,2013PhRvD..87h3008H,2019JCAP...10..029S}.
Our model implements this $H_0-N_{\rm eff}$ degeneracy, 
boosting $H_0$ in accordance with the additional radiation-like SGWB contribution,
while the coupled evolution of the Hubble parameter and tensor modes is properly taken into account.
Thus, the second part of this paper is dedicated to 
the implication of current constraints on the primordial SGWB for the Hubble tension.
We investigate the extent to which the stiff-amplified SGWB
can bring the value of $H_0$ from the CMB into agreement with those from local measurements.

The paper is organized as follows. 
In section \ref{sec:stiffamp}, we demonstrate the stiff amplification effect on the primordial SGWB 
and introduce our model.
In section \ref{sec:constraints}, we discuss all current measurements and upper bounds on the primordial SGWB, 
for each of several probes in turn.  
In section \ref{sec:combined}, we combine these probes in a joint analysis 
and derive the constraints on the ``standard inflation + stiff amplification'' scenario that result.
The implication of these results for the Hubble tension is explored in section \ref{sec:hubble}. 
We conclude in section \ref{sec:conclusion}.

\section{Stiff amplification of the primordial SGWB}\label{sec:stiffamp}

In this paper, we consider the primordial tensor perturbations with respect to a flat FLRW background metric, 
so the short-wave, weak-field limit is apparently satisfied for the GWs described by these tensor modes 
(see appendix \ref{app:GW}). 
We can write down the perturbed metric in the transverse-traceless gauge (the ``TT gauge'') 
\citep{1973grav.book.....M,1974ZhETF..67..825G}, 
$\ud s^2 = c^2\,\ud t^2 -a^2(t)(\delta_{ij}+h_{ij})\ud x^i\ud x^j$,
where $\sum_i\partial_i h_{ij}=0$ and $\sum_{i=j}h_{ij}=0$.

In section \ref{ssec:basics}, we review the basic equations concerning the primordial SGWB from inflation, 
and its amplification by a post-inflationary stiff era.
In section \ref{ssec:model}, we present our cosmological model for the ``standard inflation + stiff amplification'' scenario,
which self-consistently includes the stiff-amplified primordial SGWB.

\subsection{Basic equations}\label{ssec:basics}

Primordial tensor perturbations can be expanded in Fourier space \citep[e.g.,][]{1999PhRvD..59j2001A},
\begin{equation}
\begin{split}
    h_{ij}(t,\vec x) 
    & = \sum_{\rm P=+,\times}\int_{-\infty}^{+\infty}\ud f\int \ud^2 \hat k\,
    h^{\rm P}(t,f,\hat k)\,e^{i\vec k\cdot \vec x}\,\epsilon^{\rm P}_{ij}(\hat k),
\end{split}
\end{equation}
where $f$ is the comoving frequency, $\hat k$ is a unit vector, $\vec k\equiv 2\pi f\hat k/c$,  
and $\epsilon^{\rm P}_{ij}$ are the polarization tensors for the $+$ and $\times$ states. 
In our convention, $\sum_{i,j}\epsilon^{\rm P}_{ij}(\hat k)\epsilon^{\rm P'}_{ij}(\hat k)=2\,\delta_{\rm PP'}$.
$h^{\rm P}(t,-f,\hat k)=(h^{\rm P}(t,f,\hat k))^*$ due to the reality of $h_{ij}$.
When a mode is well-inside the Hubble radius, $h^{\rm P}(t,f,\hat k)\propto e^{-2\pi if\eta}/a$, 
where $\eta$ is the conformal time, $\ud \eta\equiv \ud t/a$.\footnote{This mode is then essentially a plane wave on time scales much shorter than the Hubble time.
It is said to satisfy the ``high-frequency'' limit (in addition to the short-wave limit) \citep{1968PhRv..166.1263I}.}

For an isotropic, stationary and Gaussian SGWB, 
the most straightforward observable is the two-point correlation function. 
In Fourier space, it is defined as
\begin{equation}
    \langle(h^{\rm P}(t,f,\hat k))^* \,h^{\rm P'}(t,f',\hat k')\rangle
    \equiv \frac{1}{2}S_h(f)\frac{\delta_{\rm D}(f-f')}{2}\frac{\delta^{\mathcal{S}^2}_{\rm D}(\hat k-\hat k')}{4\pi}\delta_{\rm PP'},
\end{equation}
where $\delta^{\mathcal{S}^2}_{\rm D}$ is the Dirac function on the two-sphere and
$S_h(f)$ is the one-sided power spectral density of the SGWB \citep{1993PhRvD..48.2389F}.
$S_h(f)$ is related to the characteristic amplitude/strain of the SGWB, $h_c(f)$,
and the (dimensionless) tensor power spectrum, $\Delta^2_h(f)$, 
by $fS_h(f)=h_c^2(f) = \Delta^2_h(f)/2$ \citep[e.g.,][]{2008PhRvD..77f3504B}.

The primordial SGWB is characterized by its power spectrum at an initial time, 
$\Delta^2_{h, \rm i}(f)\equiv\Delta^2_h(t_{\rm i},f)$,
and the tensor transfer function, $T_h(t,f) \equiv h^{\rm P}(t,f,\hat k)/h^{\rm P}(t_{\rm i},f,\hat k)$.
Standard single-field, slow-roll inflation predicts a nearly scale-invariant initial power spectrum for tensor modes, 
$\Delta^2_{h, \rm i}(f)=A_{\rm t}(f/f_*)^{n_{\rm t}}$. 
Here $A_{\rm t}$ is the tensor amplitude, 
$r\equiv A_{\rm t}/A_{\rm s}$ defines the tensor-to-scalar ratio,
and the tensor spectral index satisfies the consistency relation, $n_{\rm t} = -r/8$.
Following the convention of \emph{Planck}, the pivot scale is chosen as 
$k_* = 2\pi f_*/c = 0.05$ Mpc$^{-1}$ \citep{2020A&A...641A..10P}.
As for the transfer function, its evolution follows from the wave equation (\ref{eq:Thwe}), 
\begin{equation}\label{eq:transfer}
	\ddot T_h+\frac{3\dot a}{a}\dot T_h+\left(\frac{2\pi f}{a}\right)^2 T_h=0,
\end{equation}
where the overdot denotes the derivative with respect to the cosmic time, $t$, 
and we have omitted the arguments $(t,f)$ for brevity. 
For any mode that has undergone inflation, its amplitude is frozen while it is well-outside the Hubble radius,
so that $\Delta^2_h(t,f)\simeq \Delta^2_{h, \rm i}(f)$, $T_h\simeq1$ and $\dot T_h \simeq 0$.
After Hubble reentry, the transfer function for all modes asymptotically evolves as $T_h\propto 1/a$ (the adiabatic solution).
However, their relative amplitudes (frequency dependence) at a given time is subject to 
the EoS of the Universe at the reentry of each mode, allowing for \emph{stiff amplification}.
We will recap this effect in what follows.

For a given mode of frequency $f$, its sub-Hubble solution for $\Omega_{\rm GW}(f)$ can be approximated as 
(using eq.~[\ref{eq:subhubbleGW}])
\begin{equation}\label{eq:stiffamp}
    \Omega_{\rm GW}(a,f) \simeq \frac{(2\pi f)^2\,\Delta^2_{h, \rm i}(f)\,T_h^2}{12\,a^2H^2}
    \propto \Delta^2_{h, \rm i}(f)\left(\frac{2\pi f}{aH}\right)^2\left(\frac{a_f}{a}\right)^2,
\end{equation}
where $H\equiv \dot a/a$ is the Hubble parameter and 
$a_f\,H(a_f)\equiv 2\pi f$ defines the scale factor at its Hubble reentry (cf. eq.~[58] in LSR17).
As mentioned in the introduction, modes that reentered during the RD era 
correspond to a plateau in $\Omega_{\rm GW}(f)$ for standard inflation.

Alternatively, an early stiff era gives rise to a blue-tilted spectral shape in $\Omega_{\rm GW}(f)$. 
Such a stiff era is proposed by a variety of physical mechanisms 
\citep[e.g.,][]{1972MNRAS.160P...1Z,1982NuPhB.208..501B}
and many of them involve a scalar field dominated by its kinetic energy
\citep{1990PhRvD..42.3310K,1993PhLB..315...40S,1997PhRvD..55.1875J,1998PhRvD..57.6022J,1999PhRvD..59f3505P,2007PhRvD..76j3530C}.
In LSR17, the stiff era is due to the stiff phase of SFDM, interposed between reheating and the RD era.
We illustrate this stiff phase by an example $\Lambda$SFDM universe in appendix \ref{app:SFDM}.
For a mode that reentered during the stiff era, we showed in section III.B.3 of LSR17 that
its Hubble reentry happens later than it would if the Universe were RD all the time.
In other words, $a_{f,\rm stiff}>a_{f,\rm rad}$ for the value of $a_f$ appearing in eq.~(\ref{eq:stiffamp}).
Therefore, eq.~(\ref{eq:stiffamp}) shows that for such a mode,
the value of $\Omega_{\rm GW}(a,f)$ in the sub-Hubble limit ($a\gg a_f$)
is greater than it would be if the Hubble reentry happened in the RD era (i.e., the plateau value).
In this way, the primordial SGWB is amplified for modes reenter during the stiff era,
relative to the plateau.

\subsection{Stiff-amplified SGWB: self-consistent model for precision cosmology}\label{ssec:model}

Stiff amplification causes a secondary blue tilt in $\Omega_{\rm GW}(a,f)$ evaluated at late times 
when all modes of interest are in the sub-Hubble limit. 
Whereas any pre-RD era with an EoS stiffer than radiation would generically lead to a blue tilt 
(whose spectral index depends on the EoS),
we will only consider a stiff era ($w=1$) in this paper, in order to maximize the amplification.
Then, for modes that reentered during the stiff era, $\Omega_{\rm GW}(f)\propto f$,
$h_c(f)\propto f^{-1/2}$ (see eq.~[\ref{eq:subhubbleGWhc}]).
On the other hand, an extended period of reheating, with a matter-like EoS $(w=0)$, precedes the stiff era, as mentioned above. 
For modes that reentered during reheating, $\Omega_{\rm GW}(f)\propto f^{-2}$.
Therefore, in the ``standard inflation + stiff amplification'' scenario, the combined effect of reheating and the stiff era 
introduces an excess in the spectrum of $\Omega_{\rm GW}(f)$ relative to the plateau associated with the standard RD era, 
which appears as a triangle (in logarithmic scales; see LSR17).
This triangle peaks at $f_{\rm re}$, which corresponds to the mode that reentered at the end of reheating,
characterized by $T_{\rm re}$, the reheating temperature.

To account for stiff amplification, we consider a cosmological model 
which contains a stiff component ($w_{\rm s}=1$) and the primordial SGWB, 
in addition to all the base-$\Lambda$CDM components.\footnote{In this paper, we assume that neutrinos are \emph{massless},
so our base-$\Lambda$CDM model is slightly more simplified than that adopted by \emph{Planck}.
On the other hand, our model accounts for the thermal history in the early Universe, 
e.g., the processes of neutrino decoupling and electron-positron annihilation.} When reheating ends, virtually all of the energy density is assumed to go into the stiff component.
Thereafter, the energy density of the stiff component evolves as $\rho_{\rm s}\propto a^{-6}$ 
and dominates the total energy density of the early Universe between the end of reheating and the end of the stiff era. 
The latter endpoint is defined as the moment of equality 
between the energy density of the stiff component and that of the radiation components, 
parameterized by the temperature at this equality, $T_{\rm sr}$.
Therefore, apart from the base-$\Lambda$CDM parameters, our model has three parameters: $r$, $T_{\rm re}$ and $T_{\rm sr}$. 
As we shall describe below, the model requires us to solve a set of coupled,
integro-differential equations for each set of model parameters.  

To solve for tensor transfer functions, we apply the dynamical system approach.
For a given mode with comoving frequency $f$, the following dynamical variables can be defined:
\begin{equation}
	\zeta_f \equiv \ln \frac{2\pi f}{aH},\quad x_f \equiv \frac{\dot T_h}{H}, \quad y_f \equiv \frac{2\pi f}{aH}\, T_h.
\end{equation}
Apparently, $T_h = y_f/e^{\zeta_f}$.
The wave equation~(\ref{eq:transfer}) can then be rearranged into the following dynamical system:
\begin{subequations}\label{eq:SGWBds}
\begin{align}
	\zeta_f' & = \frac{3}{2}\sigma -1, \\
	x_f' & = -3x_f + \frac{3}{2}\sigma \, x_f - e^{\zeta_f}y_f, \\
	y_f' & = -y_f + \frac{3}{2}\sigma \, y_f + e^{\zeta_f} x_f, 
\end{align}
\end{subequations}
where the prime denotes the derivative with respect to the number of $e$-foldings, 
$N\equiv \ln{a}$ ($\ud N = H\,\ud t$), and
\begin{equation}\label{eq:sigma}
    \sigma \equiv -\frac{2\dot H}{3H^2} = \left(\frac{\rho+p}{\rho}\right)_{\rm tot} = 
    \frac{\sum_i(\rho_i+p_i)}{\rho_{\rm tot}}
    = \Omega_{\rm m}+\frac{4}{3}\,\Omega_{\rm r}+2\,\Omega_{\rm s} + \Omega_{\rm GW}+\Pi_{\rm GW},
\end{equation}
where $\Omega_{\rm GW}$ and $\Pi_{\rm GW}$ are defined in eqs.~(\ref{eq:ndrhopGW}) and (\ref{eq:rhopfourier}),
and $\Omega_{\rm m}$, $\Omega_{\rm r}$ and $\Omega_{\rm s}$ are the energy fractions of matter (CDM+baryons), 
radiation (photons+massless neutrinos) and the stiff component, respectively.
Apparently, $\sigma$ is related to the EoS of the Universe by $\sigma=1+w$.
Therefore, the evolution of each tensor mode is coupled to the expansion history of the background Universe via $\sigma$.

\begin{table}[tbp]
\hspace*{-1em}
\begin{tabular}{|c|c|c|c|c|c|c|c|}
\hline 
 & I & II & III & IV & V & VI {\scriptsize ($\Lambda$CDM)} & VII {\scriptsize ($\Lambda$CDM)} \\
\hline
& & & & & & & \\[-1em]
$r$ & 0.05 & 0.05 & 0.05 & 0.05 & 0.05 & 0.05 & 0.05 \\
$T_{\rm re}$~(GeV) & 400 & $10^5$ & $2.5\times 10^5$ & $10^7$ & $10^7$ & $10^7$ & $10^{12}$ \\
$T_{\rm sr}$~(GeV) & $9\times10^{-3}$ & 2.2 & 2.2 & 88 & $10^4$ & N/A & N/A \\[.3em]
\hline 
& & & & & & & \\[-1em]
$\Delta N_{\rm eff,\,BBN}$ & 0.44 & 0.06 & 0.37 & 0.37 & $<10^{-4}$ & 0 & 0 \\
$\Delta N_{\rm eff,\,late}$ & 0.06 & 0.06 & 0.37 & 0.37 & $<10^{-4}$ & 0 & 0 \\
$\log_{10}\,h_c(f_{\rm yr})$ & $-17.14$ & $-18.22$ & $-18.22$ & $-18.33$ & $-18.34$ & $-18.34$ & $-18.34$ \\
$\log_{10}\,\Omega_{\rm ref, LIGO}$ & $-10.32$ & $-6.70$ & $-6.67$ & $-8.30$ & $-10.33$ & $-19.99$ & $-15.60$ \\[.3em]
\hline
\end{tabular}
\caption{\label{tab:params} Example models with different model input parameters $(r, T_{\rm re}, T_{\rm sr})$.
For each model, values of the observable quantities, 
$\left(\Delta N_{\rm eff,\,BBN},\, \Delta N_{\rm eff,\,late},\, h_c(f_{\rm yr}),\, \Omega_{\rm ref, LIGO}\right)$, 
derived from the numerical solutions to the dynamical system 
described in the text for those parameters, are listed here as well. 
These observables will be discussed in section~\ref{sec:constraints}.}
\end{table}

\begin{figure}[tbp]
\centering 
\includegraphics[width=\textwidth,origin=c,angle=0]{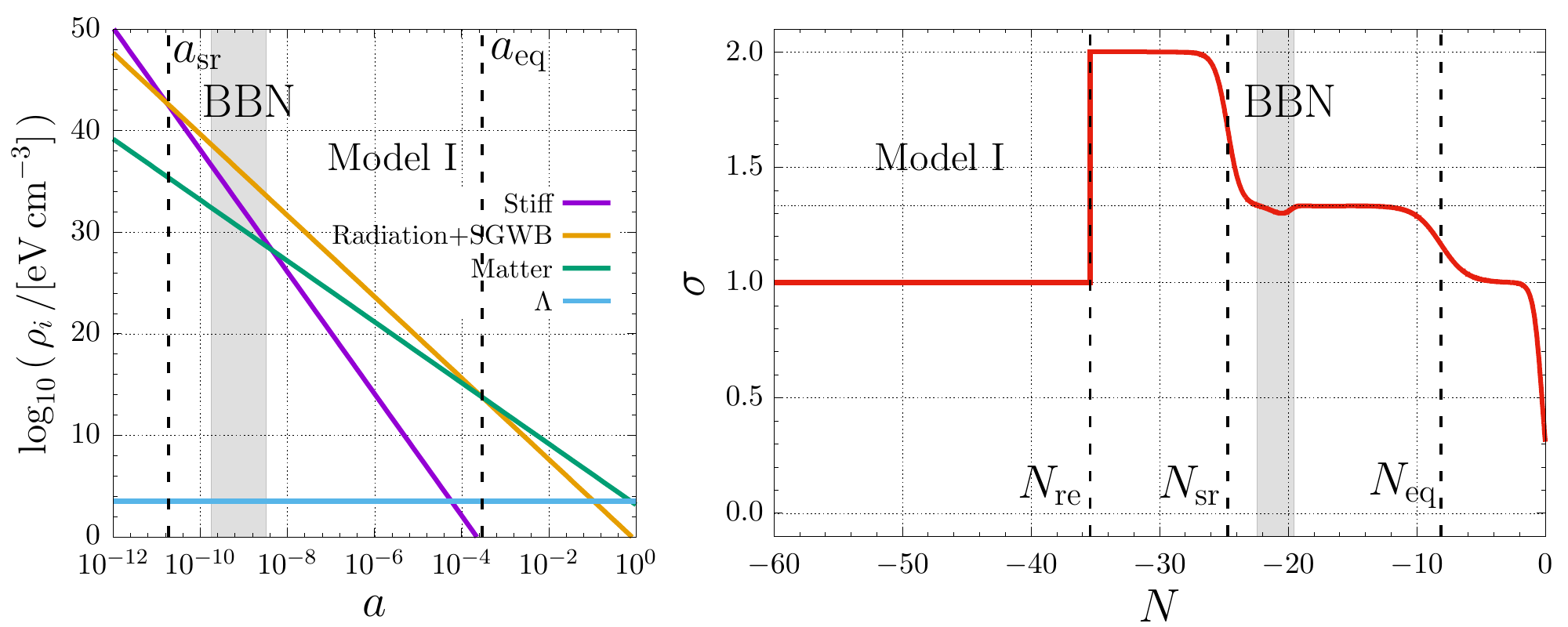}
\caption{\label{fig:model} \emph{Left panel}: Evolution of the energy density, $\rho_i$, of each component 
in our self-consistent model ($i=$ stiff, radiation+SGWB, matter, or $\Lambda$), for Model~I in table~\ref{tab:params}.
Vertical dashed lines indicate the scale factors of stiff-to-radiation and radiation-to-matter equalities, respectively.
The grey band indicates the duration of BBN.
\emph{Right panel}: Evolution of $\sigma=-2\dot H/3H^2$ for Model~I as a function of the number of $e$-foldings, $N=\ln a$.
$N_{\rm re}$ indicates the end of reheating, after which the Universe enters the stiff era.
The dip in the curve during BBN is due to the process of electron-positron annihilation.}
\end{figure}

For illustrative purposes, we have solved these equations above for several example models, 
with parameters listed in table \ref{tab:params}. 
While we will say more about our numerical method below, 
it is useful to present these examples first, 
in order to anticipate the general behavior of the solutions in the discussion which follows.
The left panel of figure~\ref{fig:model} shows the energy density evolution of each component for Model~I.
The time evolution of $\sigma$ is illustrated in the right panel of figure \ref{fig:model},
for Model~I in table~\ref{tab:params}.
In order for amplification to take place, $\sigma \neq 4/3$ is required. 
As mentioned in the introduction, when stiff amplification ($\sigma =2$) of the primordial SGWB occurs, 
the coupling between the radiation-like SGWB and the background metric 
may cause significant backreaction from the SGWB on the Hubble parameter.

To account self-consistently for this backreaction, we must solve the coupled dynamical system of eqs.~(\ref{eq:SGWBds}) and (\ref{eq:sigma})
for each frequency,  
for any given set of model parameters, $(r, T_{\rm re}, T_{\rm sr})$.
Our method of solution is described in appendix~\ref{app:numerical}.
Ordinarily, the solution of these coupled equations would be subject to boundary conditions at the present, 
fixed by the observational values adopted for $\Omega_\text{m,0}$ and $H_0$ (where $\Omega_{\Lambda,0}=1-\Omega_\text{m,0}$ for a flat FLRW Universe). 
However, observations of the CMB and baryon acoustic oscillations (BAO) also fix the value of the redshift of radiation-matter equality, $z_\text{eq}$, 
to an exquisite precision \citep[e.g.,][]{2020A&A...641A...6P}.
Since the SGWB adds an extra radiation component to the background energy density, 
we must ensure that our solution yields the observed $z_\text{eq}$, despite this.
In so doing, we encounter the degeneracy between the value of $H_0$ measured by the CMB and BAO and the boost to the radiation energy density by the SGWB,
allowed by the requirement that $z_\text{eq}$ is fixed.\footnote{In fact, it is the value of $z_\text{eq}$, which determines the size of the sound horizon,
that the CMB and BAO data are mostly sensitive to, 
rather than $\Omega_\text{m,0}$ and $H_0$.
}
As we shall show, by the end of the stiff era, the contribution of the SGWB to the background energy density reaches an asymptotic value, 
relative to that of the other radiation components. 
This asymptotic $\rho_{\rm GW}$ (which thereafter evolves as $\rho_{\rm GW}\propto a^{-4}$) 
can be represented by a constant value of $\Delta N_{\rm eff}$, the effective number of extra relativistic species. 
$\Delta N_{\rm eff}\equiv  N_{\rm eff}- N_{\rm eff,0}$, 
where $N_{\rm eff,0}=3.046$ corresponds to three Standard Model neutrinos \citep{2005NuPhB.729..221M}.
As a result, we are able to utilize the $H_0-N_{\rm eff}$ degeneracy for which the value of $z_{\rm eq}$ is preserved, 
to determine the boundary conditions in our solutions,
as follows.
While $H_0$ and $N_{\rm eff}$ can both vary, we keep $\Omega_{\rm m,0}$ and $\Omega_{\rm r,0}+\Omega_{\rm GW,0}$ fixed, 
thus fixing $z_{\rm eq}$ and $z_{\rm m\Lambda}$ (the redshift of matter-$\Lambda$ equality).
In our model, then, the $H_0-N_{\rm eff}$ degeneracy is stated as
\begin{equation}\label{eq:degeneracy}
    \frac{H_0}{H_{0,\rm \Lambda CDM}}=\sqrt{1+\mathcal{C}\,\Delta N_{\rm eff}}, \qquad
    \mathcal{C}\equiv \frac{\frac{7}{8}\left(\frac{4}{11}\right)^{4/3}}{1+\frac{7}{8}\left(\frac{4}{11}\right)^{4/3}N_{\rm eff,0}},
\end{equation}
where $H_{0,\rm \Lambda CDM}$ is the value in the base-$\Lambda$CDM model (for which $\Delta N_{\rm eff}=0$).
Thus, our model can actually help alleviate the Hubble tension by boosting $H_0$.

\begin{figure}[tbp]
\centering 
\includegraphics[width=\textwidth,origin=c,angle=0]{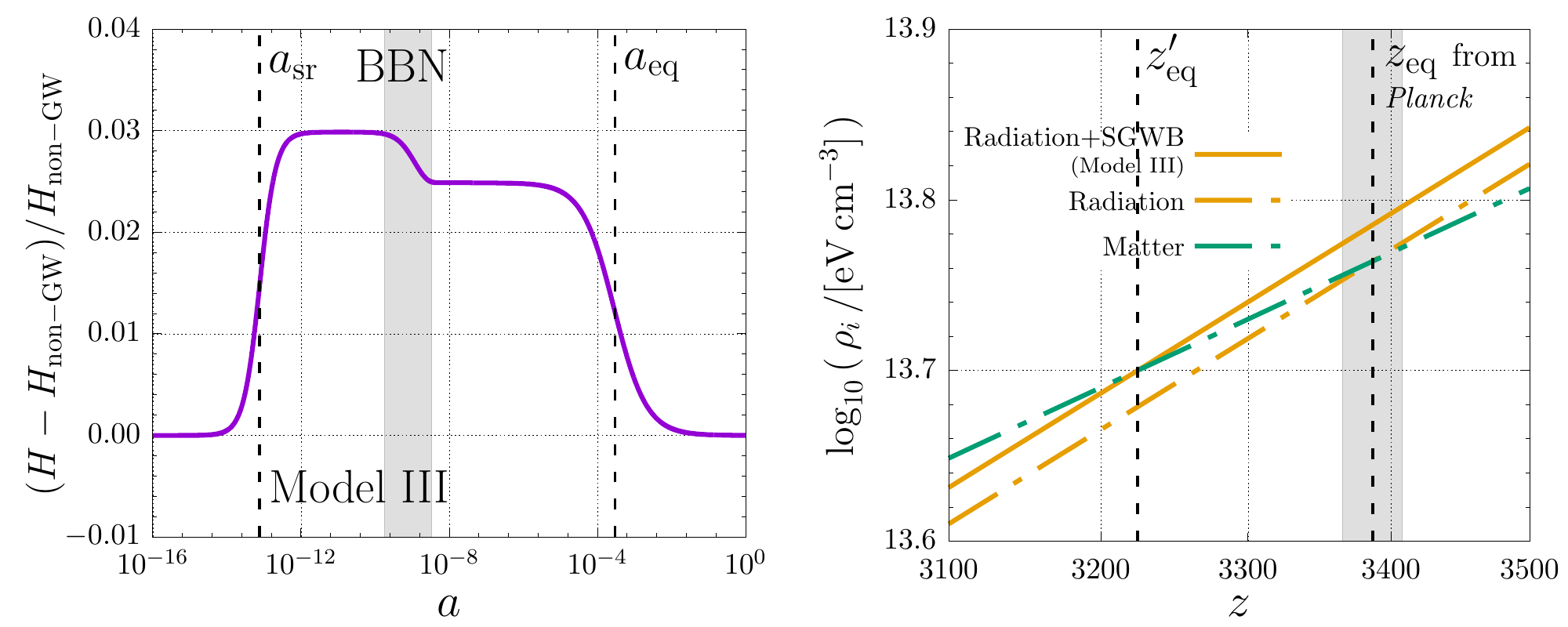}
\caption{\label{fig:compare} 
\emph{Left panel}: Fractional difference between the value of the Hubble parameter for Model~III in table~\ref{tab:params}
and that with the SGWB contribution subtracted off, 
$H^2_{\rm non-GW} \equiv H^2 - 8\pi G\,\rho_{\rm GW}/3c^2$.
The decrease of the curve during BBN is due to the process of electron-positron annihilation.
\emph{Right panel}: Shift of radiation-matter equality due to the SGWB backreaction (from $z_{\rm eq}$ to $z_{\rm eq}^\prime$), 
if $\Omega_{\rm m,0}$ and $H_0$ were both fixed at the $\Lambda$CDM best-fit values.
The solid orange line is from Model~III. 
Dash-dotted lines are from the base-$\Lambda$CDM model.
The grey band indicates the 68\% confidence interval of $z_{\rm eq}$ from \emph{Planck}'s measurements.
}
\end{figure}

Figure~\ref{fig:compare} illustrates the necessity of our treatment for the backreaction of the SGWB. 
The left panel shows that for Model~III in table~\ref{tab:params}, 
the extra radiation due to the stiff-amplified SGWB can indeed cause a $\approx3\%$ increase 
in the Hubble parameter during the RD era ($\Delta N_{\rm eff}\approx 0.37$).
The right panel shows that the backreaction of this SGWB would lead to a shift of $z_{\rm eq}$ 
more than 6$\sigma$ away from its value measured by \emph{Planck} using the base-$\Lambda$CDM model \citep{2020A&A...641A...6P},
if $\Omega_{\rm m,0}$ and $H_0$ were both fixed at the $\Lambda$CDM best-fit values.
In short, precision cosmology requires that 
the simultaneous backreaction of the primordial SGWB on the background expansion history 
be self-consistently taken into account throughout its evolution.
We have confirmed that our treatment meets this requirement with a precision $\sim 10^{-3}$ (cf. appendix~\ref{app:numerical}),
for all viable model parameters, $(r, T_{\rm re}, T_{\rm sr})$.

\begin{figure}[tbp]
\centering 
\includegraphics[width=\textwidth,origin=c,angle=0]{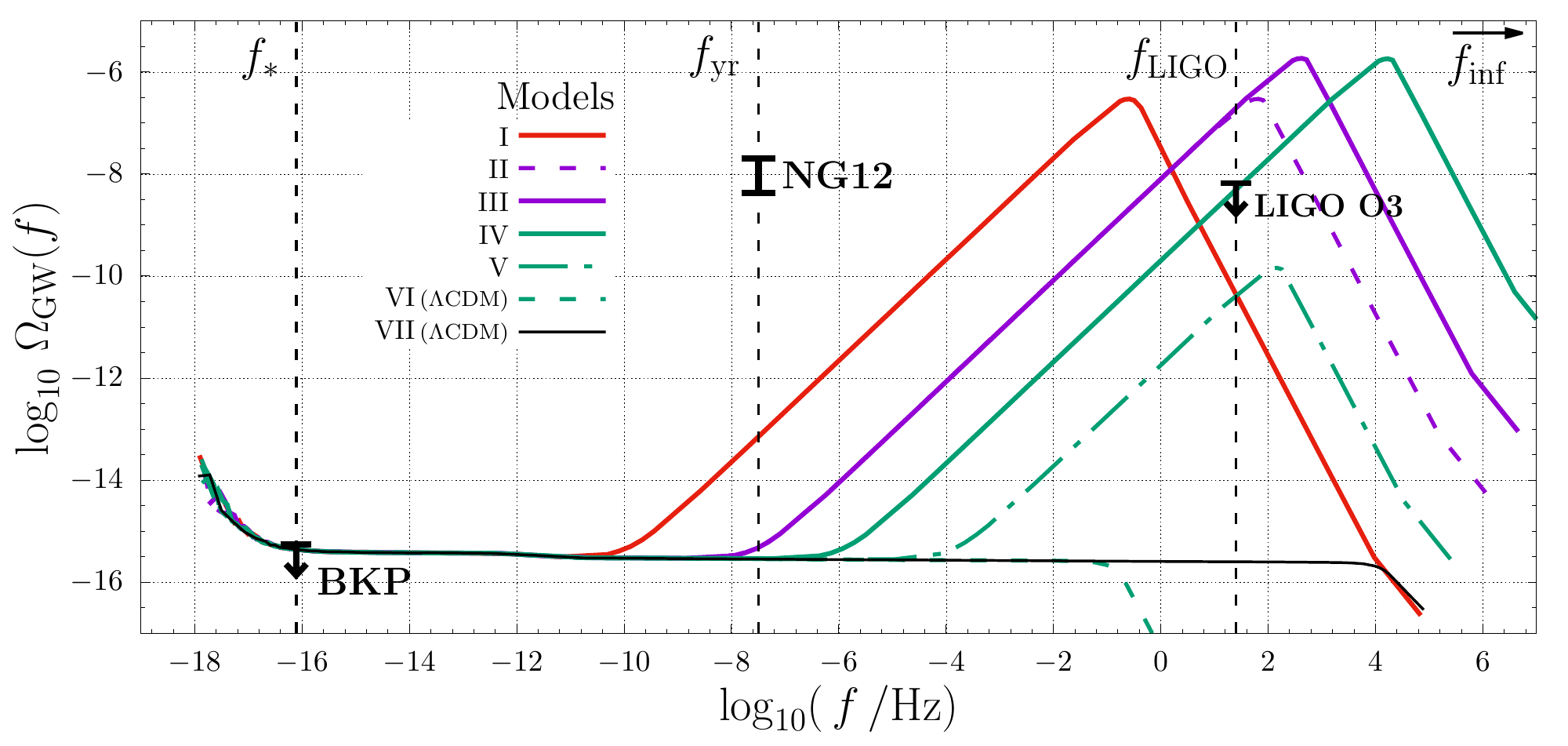}
\caption{\label{fig:OmegaGW} Present-day SGWB energy spectra for all example models in table \ref{tab:params}.
Vertical dashed lines indicate the representative frequencies of each SGWB probe, 
$f_*$, $f_{\rm yr}$ and $f_{\rm LIGO}$ for the CMB, PTA and LIGO-Virgo, respectively.
The 5\%--95\% confidence interval of the common-spectrum amplitude reported by the 12~yr NANOGrav results (labeled ``NG12'') is displayed \citep{2020ApJ...905L..34A}, 
along with the 95\% upper limit from BKP \citep{2020A&A...641A..10P} 
and that from Advanced LIGO-Virgo O3 \citep{2021PhRvD.104b2004A}.}
\end{figure}

Our results for the present-day SGWB energy spectra, $\Omega_{\rm GW}(f)$, 
for the example models in table \ref{tab:params}, are shown in figure \ref{fig:OmegaGW}.
These models are chosen so as to illustrate the dependence of the spectral shape on the model parameters.
To begin with, they all share a plateau in $\Omega_{\rm GW}(f)$ of the same height since they assume the same value of $r$.
Models~I -- V all display the blue tilt and the triangle-shaped spectrum at high frequencies 
due to stiff amplification.\footnote{Our spectral shape of $\Omega_{\rm GW}(f)$ here in figure~\ref{fig:OmegaGW}
can be compared with those in figures~8--10 in LSR17 for the $\Lambda$SFDM model,
which is a particular physical realization of our general model here and thus yields the same spectral shape for $\Omega_{\rm GW}(f)$.}
Model~I has the lowest value of $T_{\rm sr}$ and thus the highest amplitude at $f_{\rm yr} \equiv 1$~yr$^{-1}$, 
the reference frequency for PTAs.
This example also shows that when the end of the stiff era slightly overlaps BBN, 
the value of $N_{\rm eff}$ at BBN can be different from that at late times.
Models~I and II have different values of $T_{\rm re}$ and $T_{\rm sr}$ 
but the same ``area'' under the triangle in $\Omega_{\rm GW}(f)$ (as if the same triangle slides along the plateau),
which manifests itself in the equal values of $N_{\rm eff}$ for these two models at late times.
Models~II and III have the same value of $T_{\rm sr}$, so their stiff eras end at the same time.
As a result, their blue-tilted parts of $\Omega_{\rm GW}(f)$ are on top of each other, joining the plateau together.
Models~III and IV have the same values of $N_{\rm eff}$ at late times -- the highest value of all the models.
Models~IV, V and VI share the same reheating temperature, 
but their peak frequencies (at $f_{\rm re}$ for each model) are different, reflecting the different dependence of their scale factors on time.
Models~VI and VII are examples of the base-$\Lambda$CDM model. 

Throughout this paper, we adopt the following cosmological parameters 
from the \emph{Planck} 2018 results (TT,TE,EE+lowE+lensing+BAO) \citep{2020A&A...641A...6P}:
$\Omega_{\rm m,0} = 0.3111$, $z_{\rm eq}=3387$, 
$H_{0,\rm \Lambda CDM}=67.66$~${\rm km\,s}^{-1}\,{\rm Mpc}^{-1}$, $A_{\rm s}=2.105\times 10^{-9}$.

\section{Current measurements and upper bounds on the primordial SGWB}\label{sec:constraints}

In this section, we present all the current measurements and upper bounds on the primordial SGWB
from direct probes (CMB, PTA, laser interferometry) and indirect probes (BBN and late-Universe cosmology).
They are altogether illustrated in figure \ref{fig:OmegaGW}.
The constraint on our model parameters, $(r, T_{\rm re}, T_{\rm sr})$, from each probe
is examined in sections~\ref{ssec:CMBpol} through \ref{ssec:integralbound}.

\subsection{CMB temperature and polarization}\label{ssec:CMBpol}

The primordial SGWB can leave an observable imprint on the CMB temperature and polarization anisotropy
\citep{1993PhRvL..71..324C,1993PhRvD..48.4613T,1997PhRvL..78.2054S}.
In particular, detection of the CMB $B$-mode polarization around $\ell\sim 100$ 
would be a convincing signature of the primordial SGWB.
Currently, BICEP2/Keck Array+\emph{Planck} (BKP) only provides an upper bound on the tensor-to-scalar ratio,
$r<0.061$ at $95\%$ confidence level (CL) \citep{2020A&A...641A..10P}.
This upper bound directly applies to our model, too, 
since the stiff era does not affect long-wavelength modes that reentered around recombination.
In the future, CMB-S4 experiments will continuously seek to measure the primordial SGWB from inflation \citep{2016arXiv161002743A}.

\subsection{NANOGrav results}\label{ssec:NANOGrav}

\begin{figure}[tbp]
\hspace*{-2.5em}
\includegraphics[width=1.1\textwidth,origin=c,angle=0]{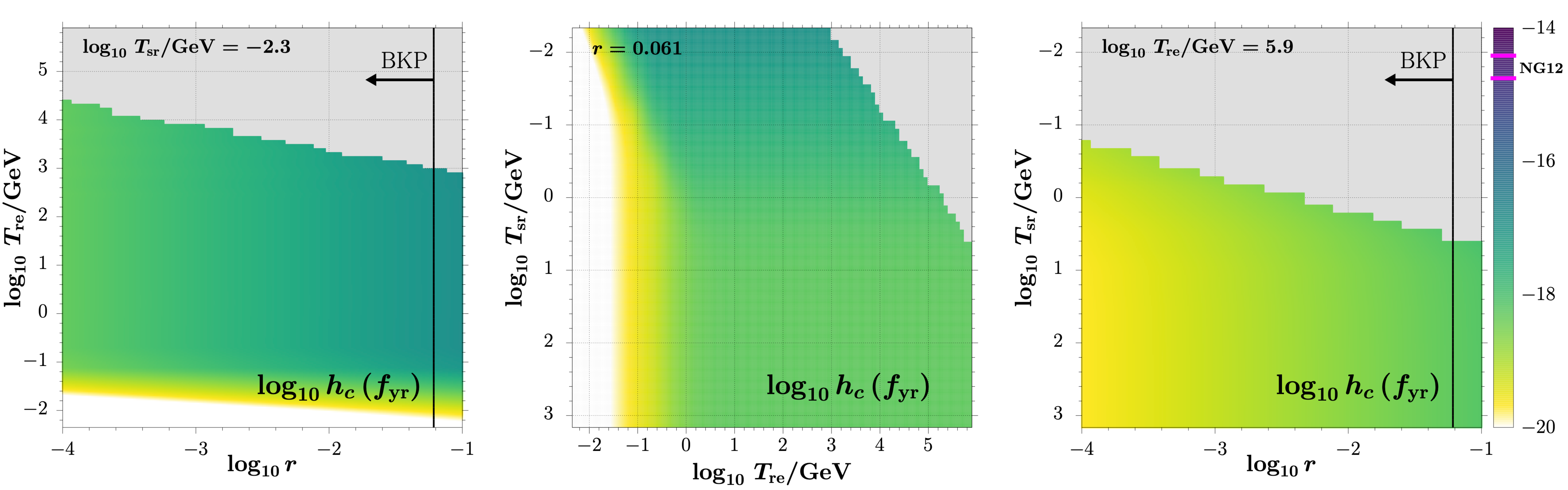}
\caption{\label{fig:hc_PTA} Characteristic strain, $h_c$, of the stiff-amplified primordial SGWB today at $f_{\rm yr}$ in our model,
presented as the three-view projections with respect to the model parameters, $(r, T_{\rm re}, T_{\rm sr})$.
The cross-sectional planar slices of the 3-D space of model parameters chosen in each view are color-coded, 
and the grey region represents parameters entirely excluded by the observational constraints. 
The vertical black line indicates the 95\% CL upper limit on $r$ from BKP \citep{2020A&A...641A..10P}.
The magenta bars on the color box (labeled by ``NG12'') indicate the 5\%--95\% confidence interval 
of the common-spectrum amplitude, $h_c(f_{\rm yr})$, from the 12.5~yr NANOGrav results \citep{2020ApJ...905L..34A}.}
\end{figure}

PTA observations measure the times of arrival (``ToAs'') of radio pulses from millisecond pulsars.
Those ToAs can be modulated by an SGWB permeating the spacetime between the pulsar and the earth.
In fact, the existence of an SGWB would be manifested in the timing-residual cross-power spectral density 
(cf. eq.~[2] in \citep{2020ApJ...905L..34A})
as a time-correlated, common-spectrum stochastic process across all pulsar-earth pairs,
with quadrupolar spatial correlations between pulsars (i.e., the Hellings \& Downs curve \citep{1983ApJ...265L..39H}). 
In PTA analysis, the characteristic strain of an SGWB is usually modeled as a power law,
$h_c(f)=A_{\rm CP}\,(f/f_{\rm yr})^\alpha$.

NANOGrav recently discovered a time-correlated, stochastic process 
with a common amplitude and spectral index in their 12.5~yr data set.
However, there is little evidence for quadrupolar spatial correlations in this common-spectrum process, 
required to identify it with an SGWB.
Hence, the NANOGrav results are still inconclusive with regard to GW detection,
and, in the meantime, have yet to be confirmed by other PTAs.
Nevertheless, despite its uncertainty, this reported common-spectrum process has incited many attempts to explain it in terms of the SGWB.

In our model, the present amplitude of $h_c$, or, equivalently, $\Omega_{\rm GW}(f)$, at frequencies near $f_{\rm yr}$ 
can be higher than in the $\Lambda$CDM model \emph{only if} the corresponding modes have experienced stiff amplification.
For example, as shown in figure~\ref{fig:OmegaGW}, these modes lie within the blue-tilted part of the SGWB spectrum for Model~I,  
so their amplitudes at $f_{\rm yr}$ are higher than the $\Lambda$CDM-plateau value.
Here, we sample our model parameters, $(r, T_{\rm re}, T_{\rm sr})$, throughout the entire parameter space,
to calculate the value of $h_c(f_{\rm yr})$ of the primordial SGWB for all model parameters of interest.
Our results are shown in figure~\ref{fig:hc_PTA} 
(with the $T_{\rm sr}$ axis upside-down in all figures in this paper that show the results for different models, spanning the range of parameter space).\footnote{Throughout our analysis, 
we do not sample the grey region in parameter space displayed in figures~\ref{fig:hc_PTA} -- \ref{fig:Delta_Neff} for computational efficiency, 
because models in this region result in too much extra radiation energy density from the stiff-amplified SGWB, 
and are thus firmly excluded by late-Universe $N_{\rm eff}$ bounds (cf. section~\ref{ssec:integralbound}).}
We compare our results with the $A_{\rm CP}$ posterior reported by NANOGrav \citep{2020ApJ...905L..34A}, 
for which the 5\%--95\% confidence interval is $1.75-3.83\times10^{-15}$ 
in the case of the blue-tilted spectral slope predicted for the stiff-amplified SGWB ($\alpha=-1/2$).\footnote{These values we quote 
are different from those in the fiducial model in NANOGrav's analysis, 
because the latter assumes $\alpha=-2/3$, as expected for the SGWB from unresolved mergers of supermassive black-hole binaries.}
Figure~\ref{fig:hc_PTA} shows that the amplitude of the SGWB at $f_{\rm yr}$ 
in the ``standard inflation + stiff amplification'' scenario, as constrained by other observations, 
is too small to explain the common-spectrum process in the 12.5~yr NANOGrav data set.\footnote{Our result here that 
the amplitude of the stiff-amplified SGWB spectrum at $f_{\rm yr}$ 
is constrained to be far below the NANOGrav results 
is qualitatively consistent with the argument in \citep{2009CQGra..26d5004G}, 
based upon applying the BBN constraint (which we shall discuss in section~\ref{ssec:integralbound}) 
to limit how late the stiff era can end. 
While the model in \citep{2009CQGra..26d5004G} differs from ours 
(e.g., it posits a stiff era that immediately follows inflation, 
with no standard reheating process), 
this reflects the fact that the example GW spectra in \citep{2009CQGra..26d5004G}, computed numerically,
share the spectral feature of ours for modes whose Hubble reentry occur during the stiff era, 
with a blue tilt of $\Omega_{\rm GW}(f)\propto f$.  
}

\subsection{Advanced LIGO-Virgo}\label{ssec:LIGOVirgo}

\begin{figure}[tbp]
\hspace*{-2.5em}
\includegraphics[width=1.1\textwidth,origin=c,angle=0]{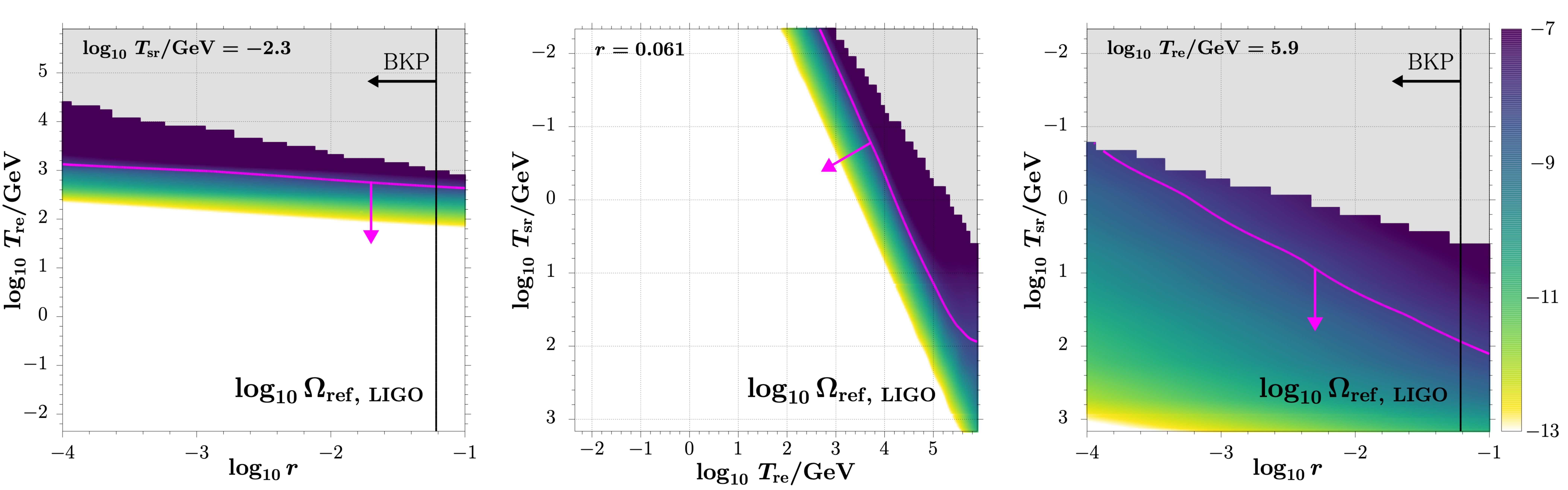}
\caption{\label{fig:Ogw_LIGO} Present-day energy density fraction per logarithmic frequency 
of the stiff-amplified primordial SGWB, $\Omega_{\rm GW}(f)$, 
at the reference frequency $f_{\rm LIGO}=25$~Hz in our model,
presented as the three-view projections with respect to the model parameters, $(r, T_{\rm re}, T_{\rm sr})$.
The cross-sectional planar slice of the 3-D space of parameters shown in each view is color-coded, and the grey region is entirely excluded by the observational constraints. 
The vertical black line indicates the 95\% CL upper limit on $r$ from BKP \citep{2020A&A...641A..10P}.
The magenta curves indicate the 95\% CL upper limit on $\Omega_{\rm ref,\,LIGO}$
from the Advanced LIGO-Virgo O3 results \citep{2021PhRvD.104b2004A}.}
\end{figure}

Laser interferometers like the Advanced LIGO-Virgo network can directly detect SGWBs 
by cross-correlating data from different detectors \citep[e.g.,][]{2017LRR....20....2R}.
Recently, the LIGO Scientific Collaboration and Virgo Collaboration published results of a search for an isotropic SGWB
using data from their first three observing runs (O1, O2 and O3) \citep{2021PhRvD.104b2004A}.
While the cross-correlation spectrum from data does not show evidence for an SGWB signal, 
a new upper limit is placed on the present-day SGWB energy spectrum, modeled as a power law,
$\Omega_{\rm GW}(f)=\Omega_{\rm ref,\,LIGO}\, (f/f_{\rm LIGO})^{\alpha_{\rm LIGO}}$.
The reference frequency is chosen to be $f_{\rm LIGO}=25$~Hz.

We again calculate the value of $\Omega_{\rm ref,\,LIGO}$ in our model, 
sampling the model parameters $(r, T_{\rm re}, T_{\rm sr})$, as shown in figure~\ref{fig:Ogw_LIGO}.
Since the stiff-amplified SGWB in our model has a triangle-shaped spectrum 
(i.e., $\Omega_{\rm GW}(f)$ is a broken power law), 
it does not always have a fixed spectral index across the LIGO-Virgo frequency range, 20--1726~Hz. 
Therefore, we compare our results with the \emph{marginalized} 95\% CL upper limit from the O3 analysis, 
$\Omega_{\rm ref,\,LIGO}<6.6\times10^{-9}$, obtained by integration over $\alpha_{\rm LIGO}$.
Figure~\ref{fig:Ogw_LIGO} displays this upper limit.

\subsection{\boldmath Integral bounds: BBN, CMB+BAO}\label{ssec:integralbound}

The primordial SGWB can also be searched by indirect probes,
e.g., light element abundances from BBN, the CMB, and large-scale structure of the Universe. 
These cosmological probes provide what is known as \emph{integral bounds} on the SGWB, 
since the observables in each probe are (indirectly) affected by the integration of $\Omega_{\rm GW}(f)$ 
over a wide range of frequencies.
In the following, we will examine all such current probes, classifying them according to 
the epoch in the expansion history of the Universe to which each probe is sensitive.

\paragraph*{Early-Universe cosmology: big bang nucleosynthesis.}

Standard BBN predicts certain relic abundances for light elements like helium-4 and deuterium
(see \citep{2016RvMP...88a5004C} for a brief review).
These abundances are sensitive to the cosmology of the background Universe during BBN (when $T\sim10^9$~K), 
in particular the baryon-to-photon ratio and the expansion rate then.
Therefore, one can infer related cosmological parameters, namely the baryon density, $\Omega_{\rm b,0}h^2$ 
(where we use $h$ here to mean the Hubble constant in units of 100~${\rm km\,s}^{-1}\,{\rm Mpc}^{-1}$),
and the effective number of relativistic species at that time, $N_{\rm eff,\,BBN}$,
by combining observations of the primordial $^4$He and D abundances with theoretical BBN calculations \citep[e.g.,][]{2008CoPhC.178..956P}.
We, henceforth, use $N_{\rm eff,\,BBN}$ to denote its value during BBN, 
in order to distinguish it from the value in the late Universe, $N_{\rm eff,\,late}$ (which affects the CMB and BAO).  
We note that, in our discussion in section \ref{ssec:model} of the asymptotic $\Delta N_{\rm eff}$ associated with $\rho_\text{GW}$, 
we were actually referring to this latter $N_{\rm eff,\,late}$. 
By contrast, the value of $N_{\rm eff,\,BBN}$ in our model 
may have contributions from \emph{both} the primordial SGWB and the stiff component, 
the latter because it increases the expansion rate of the Universe 
relative to the rate for a standard RD Universe with three neutrino species, 
even though it does not, itself, evolve like a radiation-like component.
As a result, the constraint on $N_{\rm eff,\,BBN}$ can be translated into 
constraints on the \emph{sum} of the stiff-amplified primordial SGWB and the stiff component 
(rather than on the SGWB alone) in our model, and thus on the model parameters.

In this paper, we quote the 95\% CL bounds on $N_{\rm eff,\,BBN}$, marginalized over $\Omega_{\rm b,0}h^2$,
obtained from combining measurements of the primordial $^4$He mass fraction, $Y_{\rm P}$, 
and the primordial deuterium abundance, $({\rm D/H})_{\rm P}$.
For the $Y_{\rm P}$ measurement, our baseline is from the data compilation of \citep{2015JCAP...07..011A} (A15), 
but we also quote the bounds from \citep{2014MNRAS.445..778I} (I14).
For the $({\rm D/H})_{\rm P}$ measurement, we reference the results from \citep{2014ApJ...781...31C} (C14).\footnote{We are aware of
the more recent measurements of $({\rm D/H})_{\rm P}$ \citep[e.g.,][]{2018ApJ...855..102C}.
However, we quote the result from C14 in this paper for the sake of comparison, because only this result has been combined with I14.
Moreover, the value of $N_{\rm eff,\,BBN}$ is mainly constrained by the $Y_{\rm P}$ measurement, 
only mildly dependent on $({\rm D/H})_{\rm P}$ \citep{2019JCAP...10..029S}.}
The combined observational bounds on $N_{\rm eff,\,BBN}$ are presented as follows:
\begin{subequations}\label{eq:SBBNbound}
    \begin{align}
        N_{\rm eff,\,BBN} & = 2.90\,^{+0.58}_{-0.54} \quad & (95\%,~{\rm A15 + C14}), \label{eq:Aver15}\\
        N_{\rm eff,\,BBN} & = 3.58\pm 0.40 \quad & (95\%,~{\rm I14 + C14}). \label{eq:Izotov14}
    \end{align}
\end{subequations}
The discrepancy between them is due to the moderate tension between the $Y_{\rm P}$ measurements from A15 and I14, 
which is still under debate.
It is worth noting that the lower bound from I14+C14 slightly disfavors the standard value $N_{\rm eff,0}=3.046$.

\begin{figure}[tbp]
\hspace*{-2.5em}
\includegraphics[width=1.1\textwidth,origin=c,angle=0]{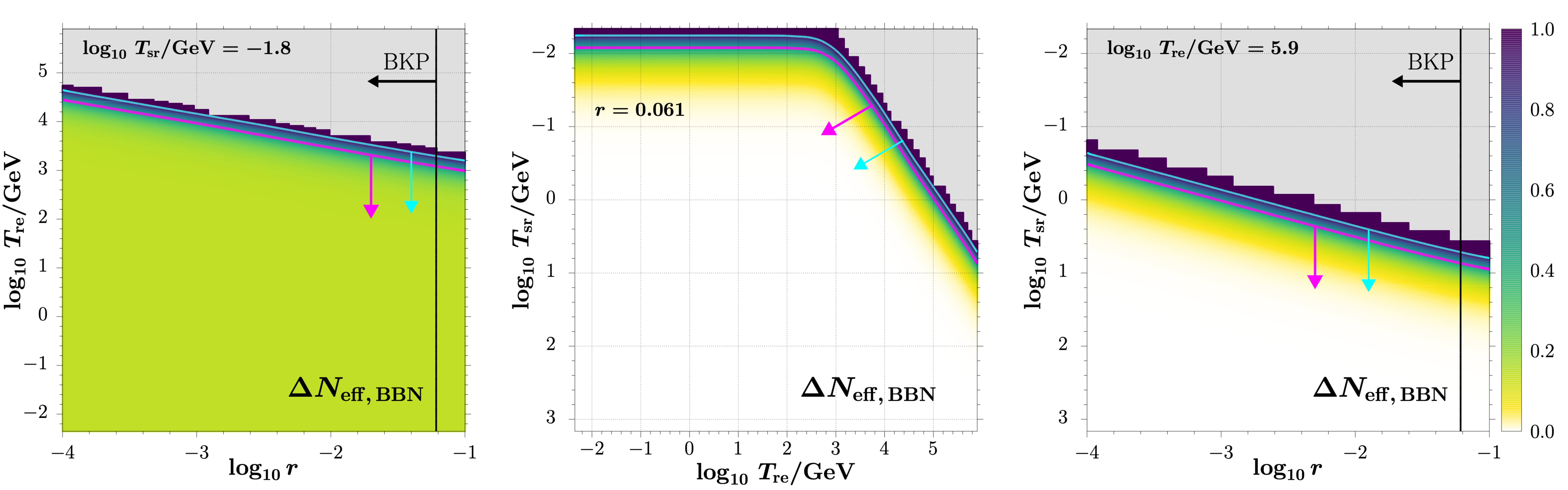}
\caption{\label{fig:DN_SBBN} Effective number of extra relativistic species during BBN, $\Delta N_{\rm eff,\,BBN}$, in our model,
presented as the three-view projections with respect to the model parameters, $(r, T_{\rm re}, T_{\rm sr})$.
The cross-sectional planar slices of the 3-D space of model parameters chosen in each view are color-coded, 
and the grey region is entirely excluded by the observational constraints. 
The vertical black line indicates the 95\% CL upper limit on $r$ from BKP \citep{2020A&A...641A..10P}.
The magenta curves indicate the 95\% CL upper limit on $\Delta N_{\rm eff,\,BBN}$
from combining the $Y_{\rm P}$ and $({\rm D/H})_{\rm P}$ measurements of A15 and C14 (eq.~[\ref{eq:Aver15}], our baseline).
The cyan curves indicate the 95\% CL upper limit from I14+C14 (eq.~[\ref{eq:Izotov14}]).}
\end{figure}

We have calculated the value of $\Delta N_{\rm eff,\,BBN}$ in our model for each choice of model parameters.  
The results are shown in figure~\ref{fig:DN_SBBN}, where the 95\% CL upper limits 
from each of the two combined observational constraints presented in eq.~(\ref{eq:SBBNbound}) are also displayed.

\paragraph*{Late-Universe cosmology: radiation-matter equality and CMB damping tail.}

\begin{figure}[tbp]
\hspace*{-2.5em}
\includegraphics[width=1.1\textwidth,origin=c,angle=0]{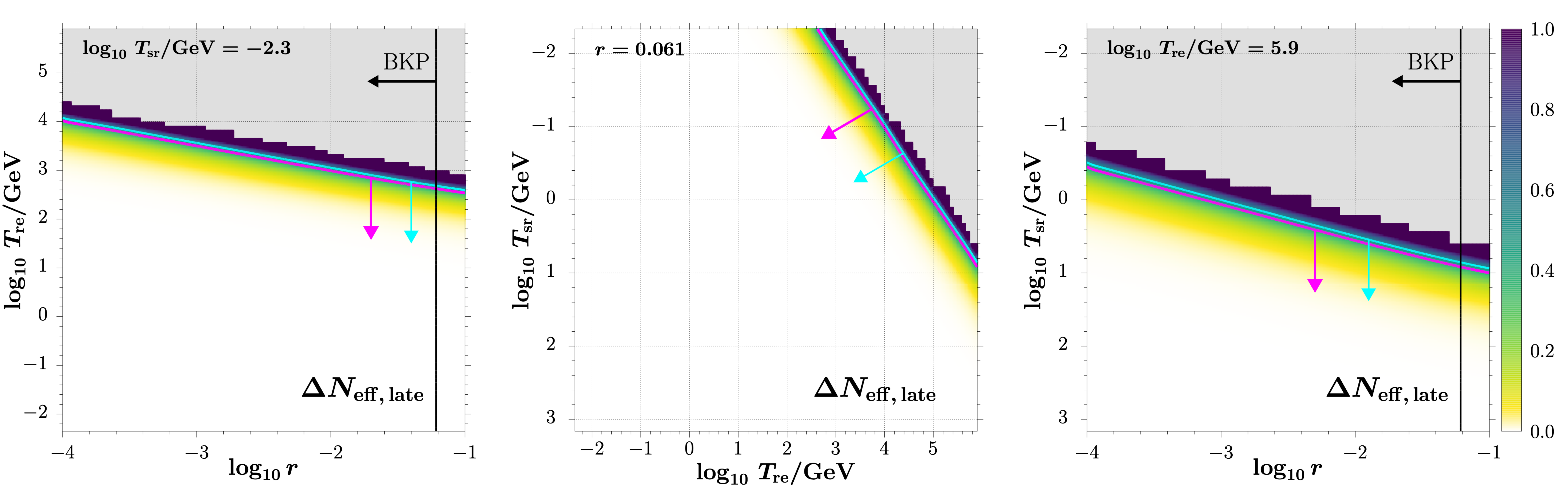}
\caption{\label{fig:Delta_Neff} Effective number of extra relativistic species in the late Universe, $\Delta N_{\rm eff,\,late}$, in our model,
presented as the three-view projections with respect to the model parameters, $(r, T_{\rm re}, T_{\rm sr})$.
The cross-sectional planar slices of the 3-D space of model parameters chosen in each view are color-coded, 
and the grey region is entirely excluded by the observational constraints. 
The vertical black line indicates the 95\% CL upper limit on $r$ from BKP \citep{2020A&A...641A..10P}.
The magenta curves indicate the 95\% CL upper limit on $\Delta N_{\rm eff,\,late}$
from combining the CMB and BAO data with a prior for $Y_{\rm P}$ (eq.~[\ref{eq:withYP}], our baseline).
The cyan curves indicate the 95\% CL upper limit from CMB+BAO only (eq.~[\ref{eq:withoutYP}]).}
\end{figure}

Extra radiation components that survive through the late Universe, 
like the stiff-amplified primordial SGWB,
may significantly contribute to the value of $N_{\rm eff,\,late}$.
In our model, unlike the case of $N_{\rm eff,\,BBN}$ discussed above, 
the SGWB is the only contribution to $\Delta N_{\rm eff,\,late}$, since the stiff component is negligible in the late Universe.
Therefore, constraints on $N_{\rm eff,\,late}$ from the CMB and BAO 
can be translated into constraints on the stiff-amplified primordial SGWB, and thus on our model parameters.
Before we discuss these late-Universe constraints,
it is important to (1) point out that a general analysis should distinguish them from the BBN constraints above 
(which are only concerned with the early-Universe cosmology),
and (2) clarify which kind of analysis is suitable for this purpose.

Currently, the CMB+BAO data sets are compatible with primordial element abundance data
inasmuch as they can be jointly fitted by an extension to the base-$\Lambda$CDM model with a fixed, 
time-independent $N_{\rm eff}$ which is common to both the early epoch of BBN and the late times probed by the CMB and BAO observations
(i.e. by assuming $\Delta N_{\rm eff,\,BBN}=\Delta N_{\rm eff,\,late}$, 
henceforth ``$\Lambda$CDM+$N_{\rm eff}$'') \citep{2020A&A...641A...6P}.
While many analyses are based on this assumption \citep[e.g.,][]{2019JCAP...10..029S},
the concordance between BBN and the late-Universe cosmology, however, 
does \emph{not} demand that these two values of $N_{\rm eff}$ be equal.
As a matter of fact, viable cosmological models like ours allow for different values of $N_{\rm eff}$ at BBN and at late times
(cf. Model~I in table~\ref{tab:params}).
In light of such models, it is therefore more general and favorable to constrain $N_{\rm eff,\,BBN}$ and $N_{\rm eff,\,late}$ separately.
For the latter, the analysis should only involve physical processes that directly determine the late-Universe observables, 
independent of BBN, in a clean way.
We carefully examine those physical processes in the following:
\begin{itemize}
    \item[(1)] The physical size of the sound horizon depends on the duration of the RD era.
    Thus, the angular scales of the sound horizon measured by both the CMB and BAO (at different redshifts)
    are sensitive to the value of $z_{\rm eq}$.
    \item[(2)] The early Integrated Sachs-Wolfe (ISW) effect refers to the enhancement of the CMB temperature anisotropies 
    due to the time-variation of gravitational potentials after recombination, 
    when the Universe was still not yet fully transitioned from RD to MD. 
    In particular, the relative heights of the first three peaks in the CMB temperature power spectrum 
    are sensitive to the value of $z_{\rm eq}$ via the early ISW effect \citep[e.g.,][]{2013ApJS..208...19H}.
    \item[(3)] On even smaller scales, the CMB temperature anisotropies are damped by photon diffusion, an effect known as Silk damping. 
    The slope of the damping tail in the CMB power spectrum reflects the amount of Silk damping \citep{2013PhRvD..87h3008H}. 
    Since it depends on both the expansion rate at recombination and the number density of free electrons,
    the CMB damping tail measurements are thus sensitive to both $N_{\rm eff,\,late}$ and $Y_{\rm P}$.
\end{itemize}

The $z_{\rm eq}$ measurements based on the first two physical processes above 
are still subject to the $H_0-N_{\rm eff}$ degeneracy, as described in section \ref{ssec:model}.
This degeneracy can, however, be broken by the CMB damping tail measurements, 
because they provide additional information that enables constraining $N_{\rm eff,\,late}$ on its own.
Therefore, one can constrain $N_{\rm eff,\,late}$ independently of information 
involving any early-Universe process (e.g., a BBN determination), 
by fitting the CMB+BAO data with an extended $\Lambda$CDM model 
which allows \emph{both} $N_{\rm eff,\,late}$ and $Y_{\rm P}$ to vary freely 
(henceforth, ``$\Lambda$CDM+$N_{\rm eff}$+$Y_{\rm P}$'').
In this paper, we quote the $N_{\rm eff,\,late}$ constraints from such an analysis 
provide by the \emph{Planck} 2018 results (CMB+BAO).\footnote{
In fact, the analysis that completely suits our purpose should use the prior on $N_{\rm eff,\,late}$ adapted for our model,
whereas the \emph{Planck} analysis was based on a conservative flat prior.
As a proof of principle, however, we quote the \emph{Planck} results in this paper. 
We leave the full analysis for our model for a future work.
}
Optionally, the $Y_{\rm P}$ value from helium abundance measurements can be additionally combined as a prior in the analysis.
This provides a tighter constraint on $N_{\rm eff,\,late}$ while still independent of BBN \citep{2020A&A...641A...6P}.
We choose this constraint as our baseline in the paper.

The $N_{\rm eff,\,late}$ constraints from \emph{Planck} are quoted as follows:
\begin{subequations}\label{eq:Nefflatebound}
    \begin{align}
        N_{\rm eff,\,late} & = 2.99\,^{+0.43}_{-0.40} \quad & (95\%,~{\rm with}~Y_{\rm P}~{\rm prior\,[A15]}), \label{eq:withYP} \\
        N_{\rm eff,\,late} & = 2.97\,^{+0.58}_{-0.54} \quad & (95\%,~{\rm without}~Y_{\rm P}~{\rm prior}). \label{eq:withoutYP}
    \end{align}
\end{subequations}
Both bounds are consistent with the standard value $N_{\rm eff,0}=3.046$.

We have calculated the value of $\Delta N_{\rm eff,\,late}$ in our model for each choice of model parameters.  
The results are shown in figure~\ref{fig:Delta_Neff},
where both 95\% CL upper limits in eq.~(\ref{eq:Nefflatebound}) are displayed.

\section{Results: joint constraints on standard inflation + stiff amplification}\label{sec:combined}

\begin{figure}[htbp]
\hspace*{-3em}
\includegraphics[width=1.1\textwidth,origin=c,angle=0]{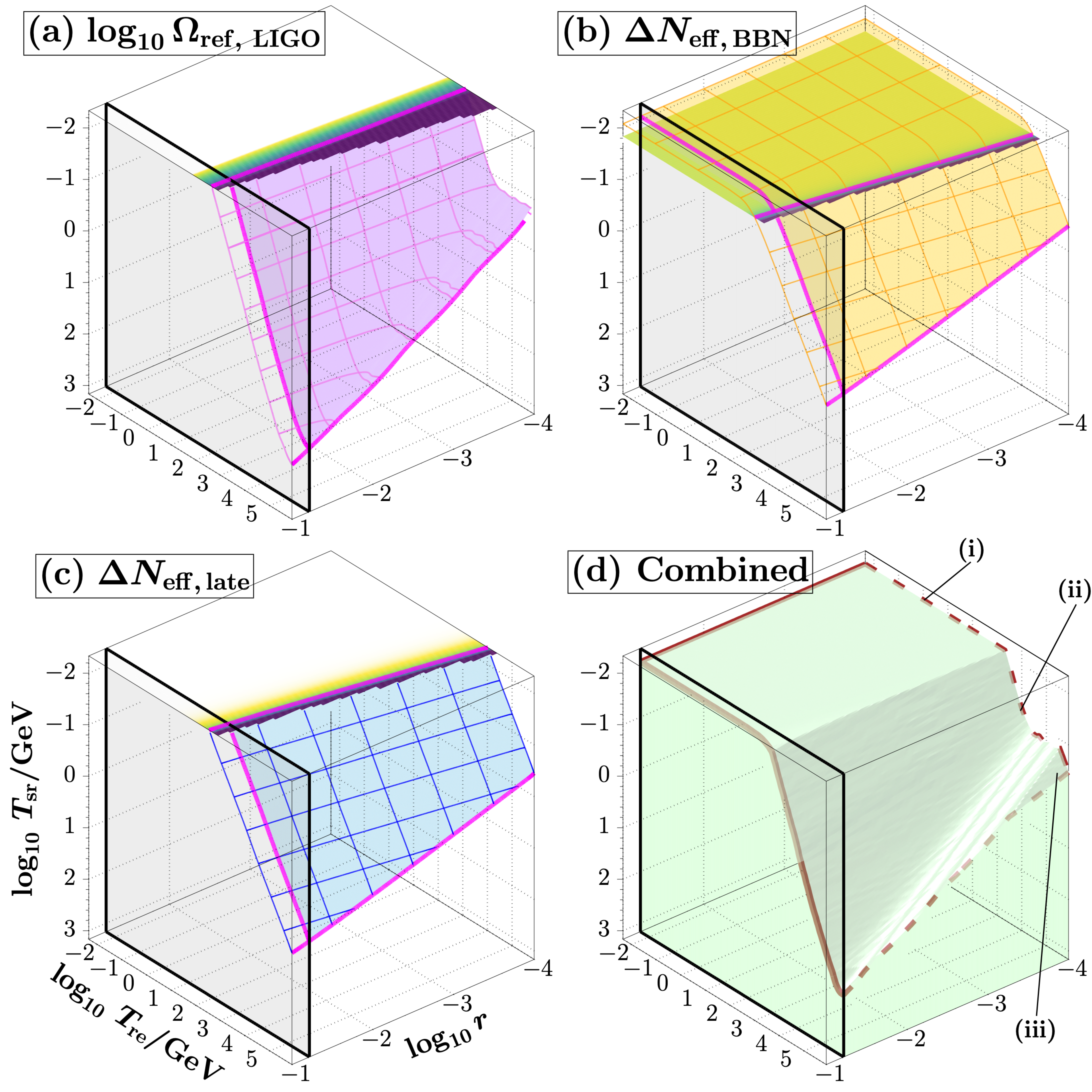}
\caption{\label{fig:3d} Three-dimensional view of the constraints on the ``standard inflation + stiff amplification'' scenario 
in its parameter space, $(r, T_{\rm re}, T_{\rm sr})$. 
\emph{Panels~(a)--(c)}: constraints from the LIGO-Virgo O3 results, 
the $N_{\rm eff,\,BBN}$ measurements and the $N_{\rm eff,\,late}$ measurements, respectively.
For each probe, the constraints are visualized as the isosurface and contours for the corresponding 95\% CL upper limit.
In panels (b) and (c), the isosurfaces are from our baseline constraints on $N_{\rm eff,\,BBN}$ and $N_{\rm eff,\,late}$, 
eqs.~(\ref{eq:Aver15}) and (\ref{eq:withYP}), respectively.
In panels~(a)--(c), the color-coded planar cross-sections are the same as those 
in the leftmost panels of figures~\ref{fig:Ogw_LIGO}--\ref{fig:Delta_Neff}, respectively.
In each panel, the three thick magenta curves are the same as those magenta curves 
shown in the three views of the corresponding figure (among figures~\ref{fig:Ogw_LIGO}--\ref{fig:Delta_Neff}),
and the grey vertical plane (with black borders in the figure) indicates the 95\% CL upper limit on $r$ from BKP \citep{2020A&A...641A..10P}.
\emph{Panel~(d)}: Overall 95\% CL allowed range of our model parameters 
obtained by combining all the constraints in panels (a)--(c), indicated by the light green volume.
We identify the three regimes of this overall allowed range according to the dominant probe in each regime, 
(i) $N_{\rm eff,\,BBN}$, (ii) LIGO-Virgo and (iii) $N_{\rm eff,\,late}$ (cf. table~\ref{tab:result}).
The 95\% CL bound in regime~(ii) is manifested as the ``waterfall''-like surface in the figure.}
\end{figure}

In this section, we combine the constraints on our model parameters, $(r, T_{\rm re}, T_{\rm sr})$, 
from all the probes of the primordial SGWB described in the previous section, 
to obtain the \emph{joint} constraints on the ``standard inflation + stiff amplification'' scenario.
We summarize these joint constraints in figure~\ref{fig:3d}, which we shall now discuss.

As already described in section~\ref{ssec:NANOGrav} and shown by figure~\ref{fig:hc_PTA}, 
the amplitude of $h_c(f_{\rm yr})$ for the stiff-amplified SGWB in our model, 
while enhanced with respect to that \emph{without} stiff amplification, in the base-$\Lambda$CDM model, 
is still much lower than the common-spectrum amplitude from the NANOGrav 12~yr data set.  
This is due to the fact that the model parameters required to amplify the primordial SGWB at this frequency 
to the level of the NANOGrav signal would result in excessively large values of $\Delta N_{\rm eff,\,late}$, 
well above its current 95\% CL upper bound from observations.
In fact, the combined constraints from all other probes indicate that 
the difference in $h_c(f_{\rm yr})$ between our model predictions and the NANOGrav results is more than two orders of magnitude.
Therefore, the ``standard inflation + stiff amplification'' scenario cannot explain the common-spectrum process reported by NANOGrav.
If the latter were indeed due to an SGWB, astrophysical sources are more likely to be the origin.

In that case, as long as we remain below the NANOGrav results, 
our model can still be constrained by other probes of the primordial SGWB,
e.g., laser interferometric experiments and indirect probes. 
The joint constraints on our model parameters from these bounds are shown in figure~\ref{fig:3d}.
It displays three-dimensional views of the 95\% CL constraints in this parameter space, 
first as required to satisfy each observational constraint separately, 
for the constraints from the O3 data of the Advanced LIGO-Virgo, the $N_{\rm eff,\,BBN}$ measurements, 
and the $N_{\rm eff,\,late}$ measurements, respectively,\footnote{For the $N_{\rm eff,\,BBN}$ and $N_{\rm eff,\,late}$ measurements, 
we only plot the isosurfaces of the tighter, baseline constraints, 
namely, eqs.~(\ref{eq:Aver15}) and (\ref{eq:withYP}), respectively.} 
and then with a view of the range of parameters allowed by \emph{all three} of those constraints --- the 95\% CL \emph{joint} constraint.
Unfortunately, since not all the likelihood data from these measurements are publicly available, 
we cannot yet perform a full Bayesian joint analysis to obtain the posteriors for our model parameters.
Instead, our joint analysis here simply combines the 95\% CL constraints on our model parameters from each probe,
i.e., combining the isosurfaces in panels~(a)--(c) of figure~\ref{fig:3d}.
The resulting 95\% CL allowed range of $(r, T_{\rm re}, T_{\rm sr})$ is indicated by the light green volume in panel~(d).

To describe the features of this overall allowed range, 
we first note that there must be a lower bound on $T_{\rm re}$ to allow BBN to occur, 
$T_{\rm re}\gtrsim 4$~MeV, and there is an upper bound on $r$ from the CMB, $r<0.061$ (95\% CL).
For fixed $r$ and $T_{\rm re}$, a lower value of $T_{\rm sr}$ (i.e., larger $\Omega_{\rm s,0}$) 
implies longer duration of the stiff era and thus higher degree of stiff amplification of the primordial SGWB.
Therefore, there must be a lower bound on $T_{\rm sr}$ for given values of $r$ and $T_{\rm re}$.
As a matter of fact, this bound is described by the \emph{top} surface of the allowed region 
in panel~(d) of figure~\ref{fig:3d}, the 95\% CL \emph{lower} limit on $T_{\rm sr}$ 
(as a reminder, the $T_{\rm sr}$ axis is upside-down is this figure).

\begin{table}[tbp]
\hspace*{-1.8em}
\begin{tabular}{|c|c|c|c|}
\hline 
& & &  \\[-1em]
Regime & Range of $T_{\rm re}$ &  Lower limit on $T_{\rm sr}$ (95\% CL)  & Dominant probe \\[.3em]
\hline
& & &  \\[-1em]
(i) & $4\times 10^{-3}\lesssim T_{\rm re}$/GeV $\lesssim 10^3$  &  $T_{\rm sr}>8.3\times10^{-3}$ GeV & $N_{\rm eff,\,BBN}$  \\[.5em]
(ii) & $10^3\lesssim T_{\rm re}$/GeV $\lesssim 10^6$ &  
~~\begin{tabular}{@{}l@{}} {\small Indicated by the ``waterfall'' surface} \\ {\small in panel~(d) of figure~\ref{fig:3d}}   \end{tabular} 
&  $\Omega_{\rm ref,\,LIGO}$ \\[1em]
(iii) & $T_{\rm re}$/GeV $\gtrsim 10^6$ 
& ~$\log_{10}\frac{T_{\rm sr}}{\rm GeV} > \frac{1}{2}\log_{10}r+\log_{10}\frac{T_{\rm re}}{\rm GeV}-4.4$ 
&  $N_{\rm eff,\,late}$~ \\[.3em]
\hline 
\end{tabular}
\caption{\label{tab:result} Overall 95\% CL allowed range of our model parameters $(r, T_{\rm re}, T_{\rm sr})$,
described as the 95\% CL lower limit on $T_{\rm sr}$ for given values of $r$ and $T_{\rm re}$
(i.e., the top surface of the allowed region in panel~[d] of figure~\ref{fig:3d}).
The overall allowed range has $r<0.061$ (95\% CL) from BKP \citep{2020A&A...641A..10P}.}
\end{table}

Using this description, we find that the parameter range allowed by the joint constraints can be characterized 
by dividing it into three regimes, according to which observational probe of our model dominates in each regime. 
They can be roughly parameterized by the range of $T_{\rm re}$. 
The results are laid out in table~\ref{tab:result} and labeled in panel~(d) of figure~\ref{fig:3d}.
We can describe these three regimes as follows, itemized by the regime number:
\begin{itemize}
    \item[(i)] This regime has the lowest values of $T_{\rm re}$. 
    The dominant constraint on $T_{\rm sr}$ is from $N_{\rm eff,\,BBN}$ 
    (cf. the discussion on the $N_{\rm eff,\,BBN}$ constraint in section~\ref{ssec:integralbound}, 
    where we explain that in our model, not only the SGWB but also the stiff component 
    can contribute to $N_{\rm eff,\,BBN}$ significantly). 
    In fact, the lower limit on $T_{\rm sr}$ in this regime is roughly manifested as a horizontal plane in the parameter space, 
    insensitive to the values of $r$ and $T_{\rm re}$.
    It reflects the fact that in this regime, the SGWB is unimportant 
    and our model is mainly constrained by the requirement that the stiff-to-radiation transition must finish early enough, 
    so that the stiff component alone, for which $\rho_{\rm s}\propto a^{-6}$, 
    does not boost the expansion rate during BBN beyond the observational constraints.
    \item[(ii)] This regime has the intermediate range of $T_{\rm re}$.
    The dominant probe is the LIGO-Virgo measurements, since the frequency of the peak in $\Omega_{\rm GW}(f)$ in our model,
    $f_{\rm re}$ --- which corresponds to the mode that reentered the Hubble radius at the end of reheating 
    (the beginning of the stiff era) --- is around $f_{\rm LIGO}=25$~Hz then.
    The lower limit on $T_{\rm sr}$ for given values of $r$ and $T_{\rm re}$ in this regime 
    is manifested as the ``waterfall'' surface shown in panel~(d) of figure~\ref{fig:3d}.
    \item[(iii)] This regime has the highest values of $T_{\rm re}$ and the dominant constraint is from $N_{\rm eff,\,late}$, 
    which amounts to an upper bound on the area under the triangle in $\Omega_{\rm GW}(f)$ (cf. figure~\ref{fig:OmegaGW}).
    Correspondingly, the lower limit on $T_{\rm sr}$ for given $r$ and $T_{\rm re}$ in this regime 
    roughly appears as a plane in the parameter space, whose equation is specified in table~\ref{tab:result}.
\end{itemize}

By contrast with our model predictions for the PTA frequency range, which fall far short of the reported NANOGrav signal for our allowed model parameters, 
there is no such gap between our model and the current upper limits from  \emph{all other} observational probes.  
As a result, if the NANOGrav signal holds up over time and is confirmed, 
then these other probes (i.e., LIGO-Virgo, $N_{\rm eff,\,BBN}$ and $N_{\rm eff,\,late}$) 
may be more likely to detect the ``standard inflation + stiff amplification'' scenario than PTA.  
For example, for the results presented here, within the range of model parameters allowed by the joint analysis, 
any future detection by LIGO-Virgo of the primordial SGWB, consistent with its current O3 95\% CL upper limit, can be explained by our model. 
If so, then our model would provide an explanation within standard inflation which does not require an initial spectral tilt.

Each of these probes will be continually upgraded in the future for its sensitivity 
(e.g., CMB-S4 \citep{2016arXiv161002743A}, LIGO A+\footnote{\url{https://dcc.ligo.org/LIGO-T1800042/public}}),
and a comparison among those sensitivities would be required to determine 
which probe will provide the first evidence of this early-Universe scenario.\footnote{
See also \citep{2014CQGra..31v5002G,1999PhRvD..60h3511B}
for more discussions on the detectability of stiff SGWB spectra by LIGO and its upgrades.
Those authors studied similar stiff SGWB spectra from standard inflation 
and discussed the dependence of its detectability on the sensitivity of GW detectors, 
but did not calculate the case with an extended phase of reheating prior to the stiff era (as in our model).
}
If detection results for multiple probes, all consistent with the predictions of ``standard inflation + stiff amplification'', 
this would then constitute smoking-gun evidence in favor of the model.

\section{Implications for the Hubble tension}\label{sec:hubble}

New observational results continue to increase the significance of the tension 
between $H_0$ measurements from CMB+BAO and those from the nearby Universe.
This has motivated growing interest in alternatives to the base-$\Lambda$CDM model \citep[e.g.,][]{2020MNRAS.498.1420W}.
Here we examine the possibility of reconciling/alleviating the Hubble tension 
in our ``standard inflation + stiff amplification'' scenario,
by the presence of the stiff-amplified primordial SGWB as an extra radiation component.  
In the presence of this extra radiation component, current measurements of  $z_{\rm eq}$ (for a precision $\sim 0.6\%$)
then drive us to incorporate the $H_0-N_{\rm eff}$ degeneracy \citep{2019JCAP...10..029S} in our analysis, 
expressed as eq.~(\ref{eq:degeneracy}) in our model.  
We note that the $H_0-N_{\rm eff}$ degeneracy is concerned with $N_{\rm eff,\,late}$,  the late-Universe value of  $N_{\rm eff}$,
independent of  $N_{\rm eff,\,BBN}$.

\begin{figure}[tbp]
\centering 
\includegraphics[width=\textwidth,origin=c,angle=0]{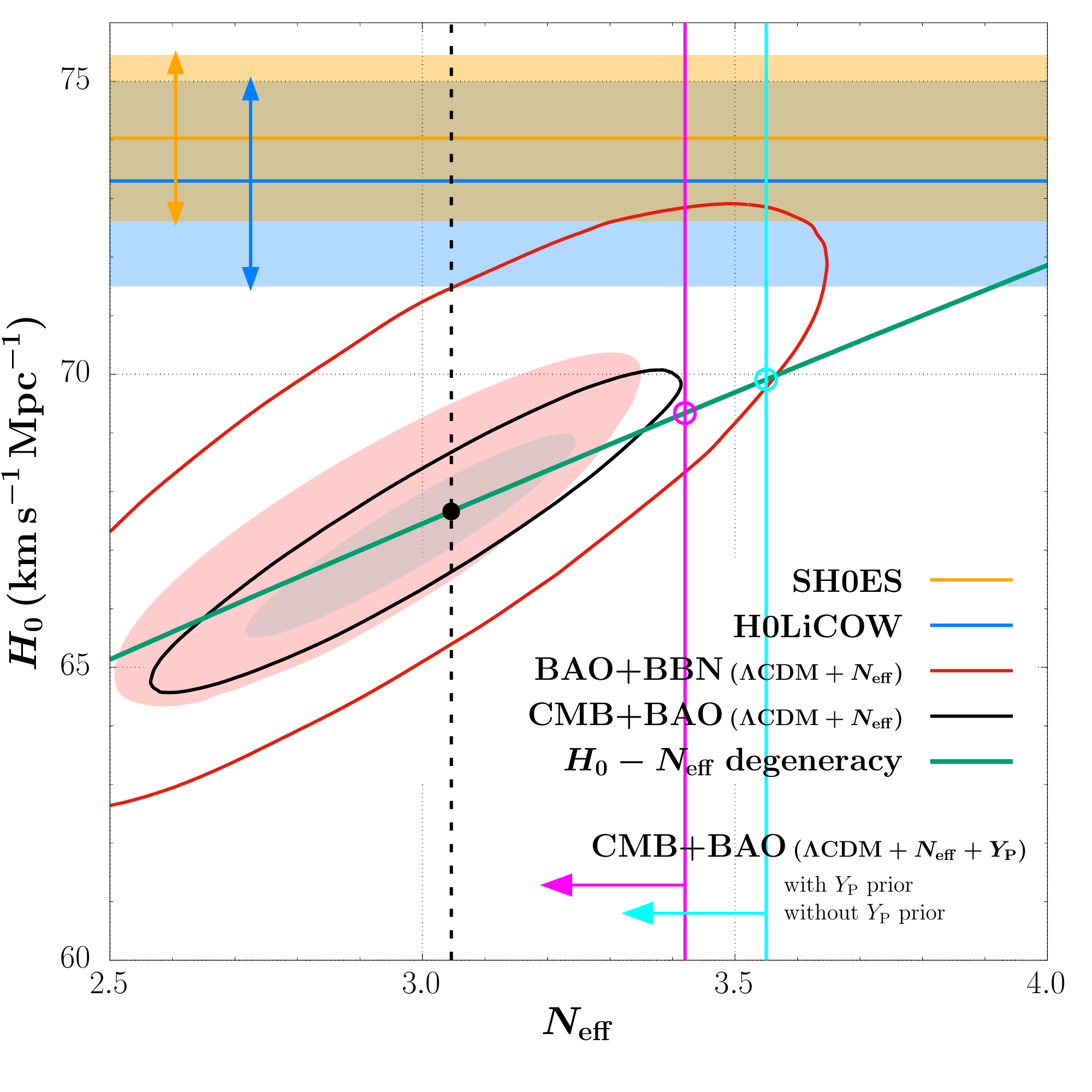}
\caption{\label{fig:hubble}
Hubble tension implications of the ``standard inflation + stiff amplification'' scenario. 
The tension is illustrated by the discrepancy between the $H_0$ measurements from the nearby Universe, 
including SH0ES \citep{2019ApJ...876...85R} and H0LiCOW \citep{2020MNRAS.498.1420W}, 
and those from the CMB and BAO, 
including BAO+BBN \citep{2019JCAP...10..029S} and CMB+BAO \citep{2020A&A...641A...6P}. 
The latter two analyses are both based on the $\Lambda$CDM+$N_{\rm eff}$ model, 
and the solid ellipses are their 95\% CL contours, respectively.
All the shaded regions represent the 68\% CL ranges for the respective measurements.
With respect to our model, the $H_0-N_{\rm eff}$ degeneracy relation, eq.~(\ref{eq:degeneracy}), is shown as the green line.
The vertical dashed line indicates the standard value, $N_{\rm eff,0}=3.046$.
The vertical magenta and cyan lines are the 95\% CL upper limits of $N_{\rm eff,\,late}$ adopted in our model, eq.~(\ref{eq:Nefflatebound}),
given by the CMB+BAO analysis based on the $\Lambda$CDM+$N_{\rm eff}$+$Y_{\rm P}$ model, with and without a prior for $Y_{\rm P}$. 
Our model thus lies on the segment of the $H_0-N_{\rm eff}$ degeneracy relation between the black point and the open circles.} 
\end{figure}

The resulting impact of our model on the Hubble tension is illustrated in figure~\ref{fig:hubble}.
Figure~\ref{fig:hubble} shows this $H_0-N_{\rm eff}$ degeneracy relation in our model, 
along with current observational determinations of $H_0$, from measurements of the distance ladder by the SH0ES collaboration 
($H_0=74.03\pm1.42$~km\,s$^{-1}$\,km$^{-1}$) \citep{2019ApJ...876...85R}, 
time-delay cosmography by the H0LiCOW collaboration
($H_0=73.3\,^{+1.7}_{-1.8}$~km\,s$^{-1}$\,km$^{-1}$) \citep{2020MNRAS.498.1420W}, 
BAO+BBN \citep{2019JCAP...10..029S} and CMB+BAO \citep{2020A&A...641A...6P}.
The last two measurements are both based on the ``$\Lambda$CDM+$N_{\rm eff}$'' model, 
and both involve a BBN calculation which assumes $N_{\rm eff,\,BBN}=N_{\rm eff,\,late}$.
In contrast, our analysis is free from these assumptions, as described in section~\ref{ssec:integralbound}.
Since the $H_0-N_{\rm eff}$ degeneracy is only concerned with $N_{\rm eff,\,late}$, 
we adopt the BBN-independent upper bound on $N_{\rm eff,\,late}$ 
from the CMB+BAO analysis provided by \emph{Planck} (eq.~[\ref{eq:Nefflatebound}]), 
based on the ``$\Lambda$CDM+$N_{\rm eff}$+$Y_{\rm P}$'' model \citep{2020A&A...641A...6P}.
Therefore, combining eq.~(\ref{eq:degeneracy}) with the \emph{Planck} fits,
we obtain the corresponding upper limits on $H_0$ for our model, as follows:
\begin{subequations}\label{eq:H0bound}
    \begin{align}
        H_0 & \leq 69.34~\mathrm{km\,s}^{-1}\,\mathrm{Mpc}^{-1} \quad & (95\%,~{\rm with}~Y_{\rm P}~{\rm prior\,[A15]}), \label{eq:H0withYP} \\
        H_0 & \leq 69.91~\mathrm{km\,s}^{-1}\,\mathrm{Mpc}^{-1}  \quad & (95\%,~{\rm without}~Y_{\rm P}~{\rm prior}). \label{eq:H0withoutYP}
    \end{align}
\end{subequations}
These values are indicated by the open circles in figure~\ref{fig:hubble}.
They occur for model parameters corresponding to the current 95\% CL upper limits of $N_{\rm eff,\,late}$, 
and thus to the boundary surface for regime~(iii) described above (cf. figure~\ref{fig:3d} and table~\ref{tab:result}). 

Meanwhile, for these current observational limits on $N_{\rm eff,\,late}$, 
our model may be able to reduce the discrepancy between measurements of $H_0$ by CMB+BAO and SH0ES 
from $4.4\sigma$ to $3.8\sigma$ for the baseline upper limit (eq.~[\ref{eq:withYP}]), 
and that between CMB+BAO and the H0LiCOW measurement from $3.1\sigma$ to $2.8\sigma$.
In addition, if we take the more relaxed upper limit, eq.~(\ref{eq:withoutYP}), 
our model may further bring the $H_0$ from CMB+BAO to within $3.4\sigma$ of the SH0ES measurement,
and within $2.6\sigma$ of the H0LiCOW measurement.

\section{Summary and discussion}\label{sec:conclusion}

The recent NANOGrav 12~yr results have invited many interpretation attempts involving a primordial stochastic gravitational-wave background.
All those attempts are based upon non-standard early-Universe scenarios which can predict stronger amplitudes in the PTA frequency range
than that from standard inflation + $\Lambda$CDM.  
While it is understood by many authors that such an SGWB contributes an extra radiation component to the background Universe,
which may therefore affect its expansion history, 
we here investigate the possibility that this extra radiation may then help alleviate the current Hubble tension, 
thus drawing a novel connection between gravitational waves and cosmology.

We demonstrate this by considering a cosmological model, the ``standard inflation + stiff amplification'' scenario, 
with two components added to the base-$\Lambda$CDM model: a stiff component and the primordial SGWB.
In our model, an early stiff era (with $w=1$) arises, when the stiff component dominates, between the end of an extended period of reheating and the standard RD era.
Unlike some other suggestions to explain the NANOGrav signal
by postulating nonstandard inflation with a blue-tilted primordial spectrum for the tensor modes responsible for the SGWB, our model does not require us to \emph{depart} from standard inflation so as to \emph{tilt} the primordial spectrum.  In our case, 
the primordial spectrum is nearly untilted, but a \emph{secondary} blue tilt results, instead, from the \emph{stiff amplification} caused by the stiff era.
Clarifying its distinction from parametric amplification, we revisit this stiff-amplification effect on the primordial SGWB 
under the general scenario considered here, which has three parameters, $(r,T_{\rm re}, T_{\rm sr})$.
The secondary blue tilt in the stiff-amplified primordial SGWB is manifested in its present-day energy spectrum as $\Omega_{\rm GW}(f)\propto f$, for the frequency range of modes that reentered the Hubble radius during the stiff era.
In this paper, we address the questions of whether such a blue tilt may explain the NANOGrav results
and to what extent the stiff-amplified primordial SGWB can reduce the Hubble tension.  Along the way, we make predictions for other direct and indirect observables of the SGWB, as well.

In doing so, we develop a new method to include the \emph{backreaction} of the SGWB on the background expansion rate \emph{self-consistently}.
In fact, we point out that any GW analysis based on a model that can significantly boost the SGWB, like ours,  
must account for its backreaction on the background Universe,
in order to preserve the well-measured redshift of radiation-matter equality as precision cosmology demands. 
For that, we solve the fully-coupled dynamical system of the SGWB and expansion history.
We update its boundary conditions by boosting the Hubble constant in accordance with the extra radiation associated with the SGWB.
In so doing, we utilize the $H_0-N_{\rm eff}$ degeneracy which preserves $z_{\rm eq}$.

We then sample the three-dimensional parameter space, $(r,T_{\rm re}, T_{\rm sr})$,
to perform a joint analysis of the NANOGrav results and the latest upper bounds from \emph{Planck}, 
big bang nucleosynthesis and Advanced LIGO-Virgo, to constrain the model.
We find that the resulting blue-tilted, stiff-amplified SGWB is still too small 
to explain the common-spectrum amplitude reported by NANOGrav (by at least two orders of magnitude),
when constrained by current upper limits of the other observables: $\Omega_{\rm ref,\,LIGO}$, $N_{\rm eff,\,BBN}$ and $N_{\rm eff,\,late}$. 
The latter together provide joint constraints on our model parameters. 
We find that the parameter range allowed by the joint constraints can be characterized 
by dividing it into three regimes, according to which observational probe of our model dominates in each regime.

While we have shown that, for its allowed parameters, 
the maximum amplitude of the predicted primordial SGWB 
for the ``standard inflation + stiff amplification'' scenario in the PTA frequency range
is far smaller than the reported NANOGrav signal, 
there is no such gap between our model predictions 
and the current upper limits on the SGWB from \emph{all other} observational probes. 
In the future, therefore, even if the NANOGrav signal is confirmed 
and too large to be the \emph{primordial} SGWB predicted here,  
these other probes (i.e., LIGO-Virgo, $N_{\rm eff,\,BBN}$ and $N_{\rm eff,\,late}$) 
may still be able to detect our predicted SGWB. 
For example, for model parameters which satisfy the constraints derived here from our joint analysis, 
any future detection of the primordial SGWB by LIGO-Virgo 
that is consistent with the current O3 95\% CL upper limit can be explained by our model.
If so, then this would be an explanation within standard inflation which does not require an initial spectral tilt.

As the sensitivity of these probes increases in the future (e.g., CMB-S4, LIGO A+), 
the chances of detecting the primordial SGWB will increase, as well.  
Which probe is likely to provide the first evidence of this early-Universe scenario 
will depend upon the relative improvements among their sensitivities over time. 
If detections occur for multiple probes, each consistent with 
the predictions of ``standard inflation + stiff amplification'', 
then this would be smoking-gun evidence in favor of the model.

With regard to the Hubble tension, we have shown that 
the ``standard inflation + stiff amplification'' scenario 
may reduce the discrepancy between the measurement of $H_0$ by CMB+BAO for the baseline upper limit (eq.~[\ref{eq:withYP}]) 
and that by SH0ES, from $4.4\sigma$ to $3.8\sigma$, 
and that by H0LiCOW, from $3.1\sigma$ to $2.8\sigma$.
Moreover, according to our analysis, if we take the more relaxed upper limit, eq.~(\ref{eq:withoutYP}), instead, 
then our model can bring the value of $H_0$ derived from CMB+BAO to within $3.4\sigma$ of its value from the SH0ES measurement,
and within $2.6\sigma$ of its value from the H0LiCOW measurement.
Hence, our results demonstrate that, while existing attempts to reconcile the Hubble tension often appeal to an extra relic radiation component,
the primordial SGWB as generally present in the current cosmological paradigm 
can provide a favorable candidate for such extra radiation which may at least partially reduce the tension.
In fact, given the unknown expansion history of the Universe between the end of inflation and BBN, 
non-negligible extra radiation is a natural consequence of the primordial SGWB produced within the standard inflationary paradigm, caused by stiff amplification, 
without further new ingredients, e.g., dark photons or early dark energy.

Our results for the ``standard inflation + stiff amplification'' scenario can also be contrasted with 
those for models recently proposed to explain the NANOGrav results as the primordial SGWB from \emph{nonstandard} inflation. 
In these nonstandard inflation models, the consistency relation between $r$ and the spectral index $n_{\rm t}$ is relaxed, 
to allow the primordial spectrum to have a large \emph{initial} blue tilt \citep[e.g.,][]{2021MNRAS.502L..11V,2021JCAP...01..071K}.  
In our model, by contrast, the inflationary consistency relation is obeyed, and, therefore, the primordial spectrum is nearly flat.
We have shown that the SGWB from \emph{standard} inflation must be well below the reported NANOGrav amplitude, even after stiff amplification,
while the nonstandard inflationary models can \emph{match} the level of that amplitude only by postulating a large enough \emph{initial} blue tilt. 
As a result, the latter models often need to place limits on the impact of reheating, 
either by restricting $T_\text{re}$ to be small (i.e., $\,\lesssim10^3$~GeV) or else invoking some nonstandard process; 
otherwise, the blue tilt required to match the NANOGrav amplitude would be so large as to violate the current BBN constraints.
Since our stiff-amplified SGWB does not rise to the level of the NANOGrav results anyway, our model is not restricted to low values of $T_{\rm re}$.

\appendix
\section{Primordial SGWB: the short-wave, weak-field limit}\label{app:GW}

The classical description of GWs is based on a clean separation of perturbations from the background metric
\citep{1968PhRv..166.1263I,1968PhRv..166.1272I,Maggiore1}. 
When such a separation occurs in spatial dimensions, this is the ``short-wave'' limit: 
all the wavelengths of interest are much less than the typical curvature radius of the background.
If perturbations are additionally small (the ``weak-field'' limit), one can expand the Ricci tensor around its background value.
In this expansion, the first-order term, $R^{(1)}_{\mu\nu}$, yields the equation of motion of the GWs
(the wave equation) and the second-order term, $R^{(2)}_{\mu\nu}$, yields their effective stress-energy tensor.
The latter is defined as
\begin{equation}\label{eq:GWwe}
	T^{\rm GW}_{\mu\nu} = -\frac{c^4}{8\pi G}\bigg\langle R^{(2)}_{\mu\nu}(\gamma)
	-\frac{1}{2}\bar g_{\mu\nu}R^{(2)}(\gamma)\bigg\rangle_{\rm 3D} 
\end{equation}
where $\bar g_{\mu\nu}$ is the background metric, $R^{(2)} \equiv \bar g^{\mu\nu} R^{(2)}_{\mu\nu}$
and $\langle\dots\rangle_{\rm 3D}$ denotes the spatial average over a scale greater than all modes of interest.

Primordial tensor perturbations over a flat FLRW background 
naturally satisfy the short-wave, weak-field limit.
Thus, in the TT gauge, the wave equation for these perturbations takes the following standard form,\footnote{
We consider no sources with anisotropic stress in this paper.}
\begin{equation}\label{eq:Thwe}
	\ddot h_{ij}+\frac{3\dot a}{a}\dot h_{ij} -\frac{c^2}{a^2}\nabla^2 h_{ij}=0.
\end{equation}
Also, $T^{\rm GW}_{\mu\nu}$ turns out to be homogeneous since it is a spatially-averaged quantity by definition.
For tensor fluctuations produced by inflation, the spatial average is equal to the ensemble average $\langle\dots\rangle$ 
according to the ergodic theorem.
Thus, primordial GWs from inflation constitute a \emph{stochastic background}.

The energy density and pressure of this SGWB can be explicitly written as
\begin{equation}\label{eq:rhopGW}
\begin{split}
	\rho_{\rm GW} & \equiv T_{00}^{\rm GW} 
	= \frac{c^2}{32\pi G}\sum_{ij}\bigg\langle\frac{1}{2}(\dot h_{ij})^2 + \frac{c^2}{2a^2}(\nabla h_{ij})^2
	+\frac{4\dot a}{a}\dot h_{ij}h_{ij}\bigg\rangle, \\
	p_{\rm GW} & \equiv \frac{1}{3a^2} (T_{11}^{\rm GW} + T_{22}^{\rm GW} + T_{33}^{\rm GW}) 
	= \frac{c^2}{32\pi G}\,\frac{1}{3}\sum_{ij}\bigg\langle -\frac{5}{2}(\dot h_{ij})^2+ \frac{7c^2}{2a^2}(\nabla h_{ij})^2 \bigg\rangle.
\end{split}
\end{equation}
Therefore, moving into Fourier space, the dimensionless energy and pressure spectra can be written as
\begin{equation}\label{eq:ndrhopGW}
\begin{split}
    \Omega_{\rm GW}(a,f) & \equiv \frac{8\pi G}{3H^2c^2}\cdot \frac{\ud\, \rho_{\rm GW}}{\ud\,\ln f} =
    \frac{\Delta^2_{h, \rm i}(f)}{24H^2}\left(\dot T_h^2 +\left(\frac{2\pi f}{a}\right)^2 T_h^2+\frac{8\dot a}{a}\dot T_h T_h\right), \\
    \Pi_{\rm GW}(a,f) & \equiv \frac{8\pi G}{3H^2c^2}\cdot \frac{\ud\, p_{\rm GW}}{\ud\,\ln f} =
    \frac{\Delta^2_{h, \rm i}(f)}{72H^2}\left(-5\dot T_h^2 +7\left(\frac{2\pi f}{a}\right)^2 T_h^2 \right).
\end{split}
\end{equation}
The inverse relations are apparently
\begin{equation}\label{eq:rhopfourier}
\begin{split}
    \Omega_{\rm GW}(a) & = \int_0^{+\infty}\Omega_{\rm GW}(a,f)\,\ud\ln f, \qquad
    \rho_{\rm GW}(a) = \frac{3H^2c^2}{8\pi G}\,\Omega_{\rm GW}(a) \\
    \Pi_{\rm GW}(a) & = \int_0^{+\infty}\Pi_{\rm GW}(a,f)\,\ud\ln f, \qquad
    p_{\rm GW}(a) =\frac{3H^2c^2}{8\pi G} \,\Pi_{\rm GW}(a). 
\end{split}
\end{equation}

For modes well-inside the Hubble radius $(2\pi f/aH\gg1)$, 
the high-frequency limit is satisfied in addition to the short-wave limit.
In this case, the adiabatic solution for plane waves reads $T_h\propto\cos{( 2\pi\,if\eta)}/a$.
It oscillates much more rapidly than the Universe expands.
Thus, only time-averaged values over several oscillations, $\langle\dots\rangle_t$, can be measurable in practice.
We then have $\dot T^2_h\simeq(2\pi f/a)^2\,T_h^2$ 
(the time-averaging notation for sub-Hubble solutions is omitted throughout the paper for brevity).
This yields 
\begin{equation}\label{eq:subhubbleGW}
    \Omega_{\rm GW}(a,f)\simeq 3\Pi_{\rm GW}(a,f) \simeq \frac{\Delta^2_{h, \rm i}(f)\,\dot T^2_h}{12H^2}
    \simeq \frac{(2\pi f)^2\,\Delta^2_{h, \rm i}(f)\,T_h^2}{12\,a^2H^2},
\end{equation}
showing that sub-Hubble modes indeed evolve like radiation, $w(a,f)=1/3$.
The last equality above also implies the following relation in the sub-Hubble limit:
\begin{equation}\label{eq:subhubbleGWhc}
    \Omega_{\rm GW}(a,f)\simeq\frac{(2\pi f)^2}{12\,a^2H^2}\Delta^2_h(a,f)=\frac{2\pi^2f^2}{3\, a^2H^2} h_c^2(a,f).
\end{equation}

\section{\boldmath Illustrative example of an early stiff era: $\Lambda$SFDM universe}\label{app:SFDM}

\begin{figure}[tbp]
\hspace*{-2.5em}
\includegraphics[width=1.1\textwidth,origin=c,angle=0]{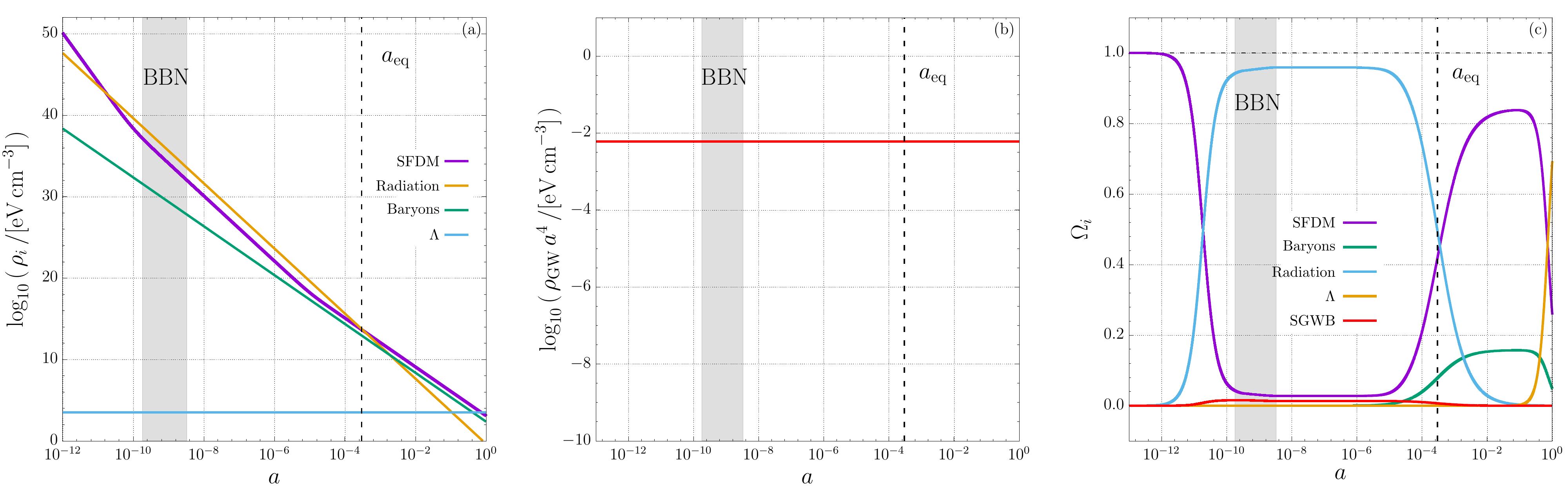}
\caption{\label{fig:LSFDM} Energy density evolution of all the components in an example $\Lambda$SFDM universe [LSR17].
Its particle parameters are $\lambda/(mc^2)^2=1 \times 10^{-18}~{\rm eV}^{-1}$cm $^3$ and $m=8\times10^{-21}$~eV$/c^2$,
and the former describes the strength of the repulsive quartic self-interaction of SFDM.
\emph{Panel~(a)}: Energy densities of SFDM, radiation, baryons and the cosmological constant.
\emph{Panel~(b)}: Energy density of the (stiff-amplified) primordial SGWB.
\emph{Panel~(c)}: Energy density fractions of all the components.
}
\end{figure}

Figure~\ref{fig:LSFDM} shows the energy density evolution of all the components in an example $\Lambda$SFDM universe, 
for which the cosmological dark matter is composed of ultralight ($m\sim 10^{-22}$~eV$/c^2$) bosonic particles 
in Bose-Einstein condensate \citep[][LSR17]{2014PhRvD..89h3536L}.
This dark matter model is described by a complex scalar field, thus known as scalar field dark matter (SFDM). 
Complex SFDM is a variant of the fuzzy dark matter (which is otherwise described by a real scalar field).
It generically undergoes a kination or stiff phase as the earliest stage of its dynamical evolution.
As a result, the primordial SGWB from inflation is subject to stiff amplification
and may then causes significant backreaction on the background Universe.

The example model shown in figure~\ref{fig:LSFDM} has a repulsive quartic self-interaction, 
which causes the radiation-like phase of SFDM as shown in panel~(a). 
The radiation-like SFDM contributes yet another extra radiation component to the critical density of the Universe, 
manifested as the corresponding plateau in panel~(c).
Later on, SFDM transitions into the matter-like phase and becomes dark matter, responsible for cosmological structure formation.

\section{\boldmath Numerical scheme: approximation by model with constant $\Delta N_{\rm eff}$}
\label{app:numerical}

As described in section~\ref{ssec:model}, 
we must solve the coupled dynamical system of eqs.~(\ref{eq:SGWBds}) and (\ref{eq:sigma}) for each frequency, 
for each set of model parameters, $(r, T_{\rm re}, T_{\rm sr})$.\footnote{For each parameter set,
we solve the dynamical system for a sample of comoving frequencies, $\{f_i\}$, 
chosen so as to resolve the spectrum $\Omega_\text{GW}(f)$ as a function of $f$, 
which in practice required about 50 frequencies, spaced more closely around the frequencies 
corresponding to the modes that entered the Hubble radius at the transitions between epochs, 
when the EoS of the Universe changed, e.g., at $T_\text{re}$.}
It is necessary, in fact, to solve the system of equations for \emph{all} frequencies at once, 
since it is a set of integro-differential equations, 
in which $\Omega_{\rm GW}$ and $\Pi_{\rm GW}$ in eq.~(\ref{eq:sigma}) are both integrals over all frequencies. 
An iterative solution is required, in that case, since the integrated quantities at a given time 
are not known until the solution is known for each frequency and at all times.
Moreover, the finite-difference scheme must contend with the requirement of 
resolving the high-frequency oscillatory behavior of the solution in time, which requires many small steps. 
Even if we only integrate the dynamical system exactly during the Hubble reentry for each mode (for about 10 $e$-foldings)
and stitch that solution with its analytical super-Hubble and sub-Hubble asymptotes,
the total solution to the system consisting of all frequencies 
can be costly for a single set of model parameters alone. 
In addition, in order to constrain our model parameters 
by comparing the solutions for different parameters with observational constraints, 
we must sample a large grid of representative points in the three-dimensional parameter space, $(r, T_{\rm re}, T_{\rm sr})$,
and find the solution to the coupled equations \emph{for each point}.

Faced with these computational challenges, 
we have developed an efficient numerical scheme to solve the coupled system.  
It takes advantage of the fact that in the exact solution for cases with significant stiff amplification,
the contribution of the SGWB to the background energy density 
reaches an asymptotic value by the end of the stiff era, 
i.e., a constant fraction of that of other radiation components.
As a result, $\rho_{\rm GW}$ can be represented in terms of a constant value of $\Delta N_{\rm eff}$.
During the stiff era, itself, the expansion history is not sensitive to this value, 
except that it determines when the stiff era ends and the Universe becomes RD. 
As such, if we knew what that asymptotic value of $\Delta N_{\rm eff}$ was going to be, 
we could approximate the entire expansion history 
and therefore the evolution of $\sigma$ (cf. eq.~[\ref{eq:sigma}]) quite well, 
by adopting this asymptotic $\Delta N_{\rm eff}$ 
and assuming that it is constant from the beginning of the integration forward in time. 
Since the evolution of $\sigma$ solely determines the transfer of tensor modes, as eq.~(\ref{eq:SGWBds}) implies,
this method would yield a good approximate solution to the coupled equations. 
Unfortunately, we do not know this asymptotic $\Delta N_{\rm eff}$ in advance of solving the coupled equations. 
However, we can approach this value iteratively, 
if we have an initial guess for the value of $\Delta N_{\rm eff}$.
In the end, this approximation enables us to produce computationally-efficient solutions,
with negligible differences from the exact solutions.

\begin{figure}[tbp]
\centering 
\includegraphics[width=\textwidth,origin=c,angle=0]{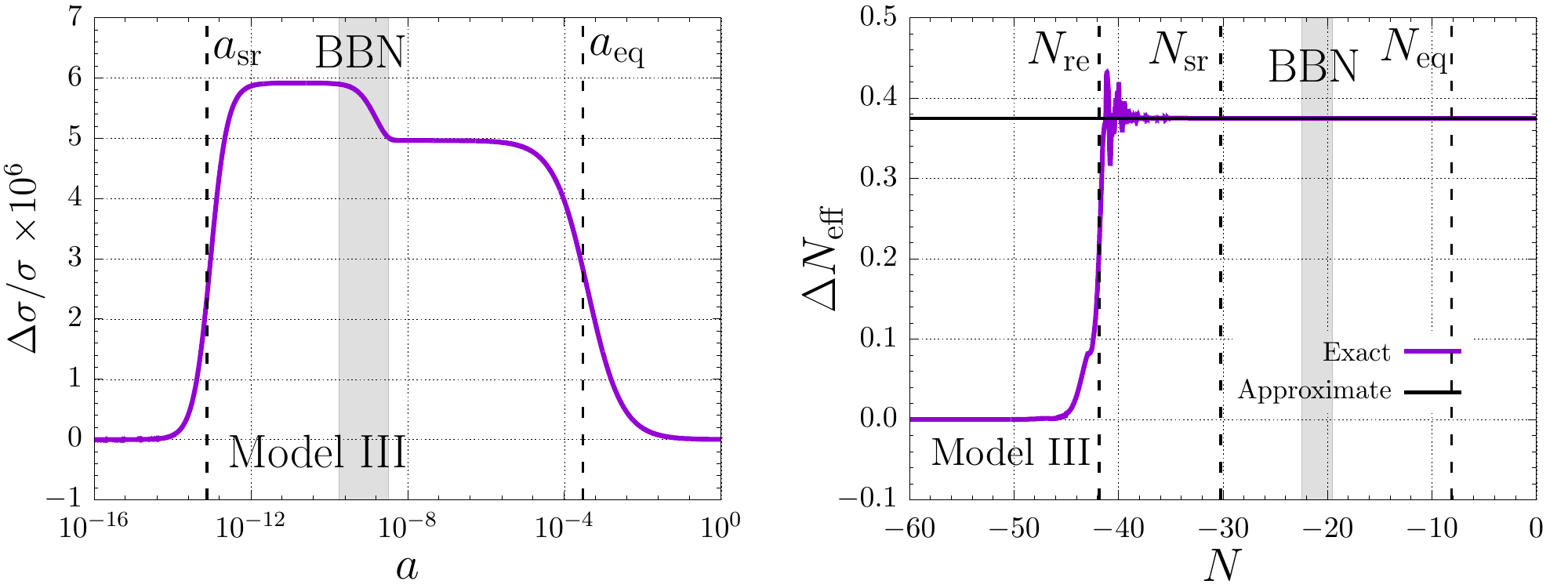}
\caption{\label{fig:approx} 
\emph{Left panel}: Fractional difference between the exact solution and the approximate model, in terms of $\sigma$, 
for Model~III in table \ref{tab:params}. 
Vertical dashed lines indicate the scale factors of stiff-to-radiation and radiation-to-matter equalities, respectively.
The decrease of the curve during BBN is due to the process of electron-positron annihilation.
\emph{Right panel}: Evolution of $\Delta N_{\rm eff}$ as a function of the number of $e$-foldings, $N$, 
for both the exact solution and the approximate model.
Vertical dashed lines indicate the end of reheating, stiff-to-radiation equality and radiation-to-matter equality, respectively.
In both panels, the grey band indicates the duration of BBN.
}
\end{figure}

In what follows, we describe this approximate model for the treatment of the backreaction of the SGWB 
(mentioned in section~\ref{ssec:model} with other detail) and justify our use of it. 
First, we explain why the value of $\Delta N_{\rm eff}$ due to $\rho_{\rm GW}$ is asymptotically constant
in cases with significant stiff amplification.
The degree of stiff amplification depends on the duration of the stiff era.
If the latter spans more than a few $e$-foldings, 
the frequency range of modes that reentered the Hubble radius during the stiff era 
will extend more than a few orders of magnitude, accordingly.
In this case, combining eq.~(\ref{eq:rhopfourier}) 
and the spectrum of the stiff-amplified SGWB ($\Omega_{\rm GW}(f)\propto f$), 
one can shown that $\rho_{\rm GW}$ must be dominated by high-frequency modes 
that reentered at the beginning of the stiff era (cf. figure~\ref{fig:OmegaGW}).
Since the tensor modes within a fixed frequency range evolve like radiation in the sub-Hubble limit (cf. eq.~[\ref{eq:subhubbleGW}]),
the overall stiff-amplified primordial SGWB can therefore be well approximated 
by a radiation component with a constant $\Delta N_{\rm eff}$, shortly after the onset of the stiff era.
As a result, we can replace eq.~(\ref{eq:sigma}) in the exact dynamical system by the following approximation:
\begin{equation}\label{eq:approxsigma}
    \sigma = -\frac{2\dot H}{3H^2} 
    = \Omega_{\rm m}+\frac{4}{3}\,\Omega_{\rm r}+2\,\Omega_{\rm s} + \frac{4}{3}\,\Omega_{\rm er},
\end{equation}
where $\Omega_{\rm er}$ is the energy fraction of this extra radiation component. 
The simultaneous coupling between the primordial SGWB and the background Universe can thus be approximated 
by the system of eqs.~(\ref{eq:SGWBds}) and (\ref{eq:approxsigma}), 
which is more efficient to solve than the exact system.

We solve this approximate system iteratively with an update on its present-day boundary conditions for each iteration. 
Particularly, we calculate the asymptotic value of $\Delta N_{\rm eff}$ associated with $\rho_{\rm GW}$ from the last iteration,
and then update the value of $H_0$ as part of the boundary conditions for the next iteration,
using the $H_0-N_{\rm eff}$ degeneracy relation, eq.~(\ref{eq:degeneracy}).
As a reminder, the sum $\Omega_{\rm r,0}+\Omega_{\rm er,0}$ that appears as a parameter in eq.~(\ref{eq:approxsigma}) 
is fixed in our treatment so as to fix $z_{\rm eq}$ (cf. section~\ref{ssec:model}). It needs no update between iterations, therefore.
We adopt the following convergence criteria for the iterative scheme that 
the fractional difference between the asymptotic values of $\Delta N_{\rm eff}$ from consecutive iterations is less than $10^{-3}$.
Fortunately, only a few iterations are required in order to converge 
or else to reach the conclusion that the adopted model parameters be excluded 
(i.e., the grey region in figures~\ref{fig:hc_PTA}--\ref{fig:Delta_Neff}).

Justification of our numerical scheme is illustrated in figure~\ref{fig:approx}, 
using Model~III in table~\ref{tab:params} as an example.
The left panel demonstrates the consistency between the exact model (with the primordial SGWB)
and the corresponding approximate model (with its converged value of the asymptotic $\Delta N_{\rm eff}\approx 0.37$), in terms of $\sigma$.
It shows that their relative difference in $\sigma$ is less than $10^{-5}$ throughout the expansion history.
The right panel displays the evolution of $\Delta N_{\rm eff}$ in both models.
It confirms that whenever the backreaction is important, that is, during the RD era,
the value of $\Delta N_{\rm eff}$ from the approximate model agrees with that from the exact model.
In summary, the self-consistent expansion history for the exact model can be faithfully mimicked 
by that from the computationally-efficient, approximate model.

\acknowledgments

BL acknowledges that this work is supported by National Key R\&D Program of China (grant No.~2018YFA0404502), 
NSFC (grant No.~11821303), and National SKA Program of China (grant No.~2020SKA0110401).
We thank Aaron Zimmerman, Kejia Lee, Xingjiang Zhu and Paulo Montero-Camacho for valuable comments and discussions,
and thank the anonymous referee for constructive suggestions.

\paragraph{Note added.} 

After we submitted the paper, several recent works are brought to our attention. 
In \citep{2021arXiv210809299C}, a triangle-shaped SGWB energy spectrum similar to ours is found. 
It is also due to the stiff amplification effect, caused by the kination phase of a scalar field.
\citep{2021arXiv210713351K} realized that a full numerical solution requires 
solving an integro-differential equation and they developed an \emph{iterative} algorithm 
with the same rationale as our solution here. 
Concerning the Hubble tension, \citep{2020PhRvD.102b3518V} also studied the possibility of 
reducing the discrepancy by extra radiation species, who derived an $H_0-N_{\rm eff}$ degeneracy relation 
different from our eq.~(\ref{eq:degeneracy}) here, based on a data-driven view. 
Other attempts to resolve the Hubble tension include \citep{2021ApJ...912..150D}, 
who suggested that current Type Ia supernovae data may imply an evolution trend 
which then reduces the tension when extrapolated to the redshift of recombination.









\bibliography{SGWB_PTA_LIGO}{}
\bibliographystyle{JHEP}

\end{document}